\documentclass[journal]{IEEEtran}

\usepackage{amsmath,amsfonts}
\usepackage{algorithmic}
\usepackage{algorithm}
\usepackage{array}
\usepackage[caption=false,font=normalsize,labelfont=bf, justification=centering]{subfig}
\usepackage{caption}
\usepackage{textcomp}
\usepackage{stfloats}
\usepackage{url}
\usepackage{verbatim}
\usepackage{graphicx}

\usepackage{cite}
\usepackage[acronyms,nonumberlist,nopostdot,nomain,nogroupskip]{glossaries}
\newacronym{mmwave}{mmWave}{millimeter wave}
\newacronym{phy}{PHY}{physical layer}
\newacronym{mac}{MAC}{medium access control}
\newacronym{uav}{UAV}{unmanned autonomous vehicle}
\newacronym{em}{EM}{electromagnetic}
\newacronym{iot}{IoT}{Internet of Things}
\newacronym{ml}{ML}{machine learning}
\newacronym{drl}{DRL}{deep reinforcement learning}
\newacronym{urc}{URC}{ultra-reliable computing}
\newacronym{urllc}{URLLC}{ultra-reliable low-latency communication}
\newacronym{miso}{MISO}{multiple-input single-output}
\newacronym{mimo}{MIMO}{multiple-input multiple-output}
\newacronym{mu}{MU}{multi-user}
\newacronym{rfid}{RFID}{Radio Frequency Identification}
\newacronym{rfp}{RFP}{radio fingerprinting}
\newacronym{sdr}{SDR}{software-defined radio}
\newacronym{mas}{MAS}{Mobile Autonomous System}
\newacronym{rl}{RL}{reinforcement learning}
\newacronym{los}{LoS}{line of sight}
\newacronym{dnn}{DNN}{deep neural network}
\newacronym{fpga}{FPGA}{field-programmable gate array}
\newacronym{cv}{CV}{computer vision}
\newacronym{mcs}{MCS}{modulation and coding scheme}
\newacronym{soc}{SoC}{system-on-chip}
\newacronym{mumimo}{MU-MIMO}{\gls{mu}-\gls{mimo}}
\newacronym{dsp}{DSP}{digital signal processing}
\newacronym{snr}{SNR}{signal-to-noise ratio}
\newacronym{csi}{CSI}{channel state information}
\newacronym{svd}{SVD}{singular value decomposition}
\newacronym{ism}{ISM}{industrial, scientific and medical}
\newacronym{dsa}{DSA}{dynamic spectrum access}
\newacronym{cnn}{CNN}{convolutional neural network}
\newacronym{rfsoc}{RFSoC}{RF System on Chip}
\newacronym{ntp}{NTP}{network time protocol}
\newacronym{ptp}{PTP}{precise time protocol}
\newacronym{nu}{NU}{Northeastern University}
\newacronym{tamu}{TAMU}{Texas A\&M University}
\newacronym{ncsu}{NCSU}{North Carolina State University}
\newacronym{uta}{UTA}{University of Texas Austin}
\newacronym{fiu}{FIU}{Florida International University}
\newacronym{uo}{OU}{University of Oklahoma}
\newacronym{gt}{GT}{Georgia Tech}
\newacronym{ucb}{UCB}{University of California Berkeley}
\newacronym{ucsb}{UCSB}{University of California Santa Barbara}
\newacronym{ttu}{TTU}{Texas Tech University}
\newacronym{uh}{UH}{University of Hawaii}
\newacronym{eess}{EESS}{Earth exploration satellite services}
\newacronym{nsf}{NSF}{National Science Foundation}
\newacronym{ntia}{NTIA}{National Telecommunications and Information Administration}
\newacronym{rfc}{RFC}{Request for Comments}
\newacronym{dod}{DoD}{Department of Defense}
\newacronym{frp}{FRP}{foundational research principle}
\newacronym{fec}{FEC}{forward error and erasure correction coding}
\newacronym{cct}{CCT}{technological cross-cutting theme}
\newacronym{wiot}{WIoT}{Institute for the Wireless Internet of Things}
\newacronym{ewd}{EWD}{education and workforce development}
\newacronym{pawr}{PAWR}{Platforms for Advanced Wireless Research}
\newacronym{ppo}{PPO}{\gls{pawr} Project Office}
\newacronym{iq}{IQ}{in-phase and quadrature}
\newacronym{if}{IF}{intermediate}
\newacronym{pl}{PL}{physical logic}
\newacronym{lna}{LNA}{low-noise amplifier}
\newacronym{rf}{RF}{radio frequency}
\newacronym{wpan}{WPAN}{wireless personal area networks}
\newacronym{wlan}{WLAN}{wireless local area networks}
\newacronym{wan}{WAN}{Wide Area Networks}
\newacronym{6g}{6G}{sixth generation}
\newacronym{vr}{VR}{virtual reality}
\newacronym{ar}{AR}{augmented reality}
\newacronym{src}{SRC}{Semiconductor Research Corporation}
\newacronym{darpa}{DARPA}{Defense Advanced Research Projects Agency}
\newacronym{adc}{ADC}{analog to digital converter}
\newacronym{dac}{DAC}{digital to analog converter}
\newacronym{gsps}{GSps}{Gigasamples-per-second}
\newacronym{awg}{AWG}{arbitrary waveform generator}
\newacronym{dso}{DSO}{digital storage oscilloscope}
\newacronym{nlos}{NLoS}{non line of sight}
\newacronym{thz}{THz}{terahertz}
\newacronym{si}{Si}{silicon}
\newacronym{soi}{SoI}{Silicon-on-Insulator}
\newacronym{sige}{SiGe}{Silicon-Germanium}
\newacronym{inp}{InP}{Indium Phosphide}
\newacronym{gan}{GaN}{Gallium Nitride}
\newacronym{gaas}{GaAs}{Gallium Arsenide}
\newacronym{jpl}{JPL}{Jet Propulsion Laboratory}
\newacronym{ic}{IC}{integrated circuit}
\newacronym{hbt}{HBT}{heterojunction bipolar transistor}
\newacronym{hemt}{HEMT}{high-electron mobility transistor}
\newacronym{pa}{PA}{power amplifier}
\newacronym{hdl}{HDL}{hardware description language}
\newacronym{fft}{FFT}{fast Fourier transform}
\newacronym{css}{CSS}{chirp spread spectrum}
\newacronym{dsss}{DSSS}{direct-sequence spread spectrum}
\newacronym{rssi}{RSSI}{received signal strength indicator}
\newacronym{bs}{BS}{base station}
\newacronym{ue}{UE}{user equipment}
\newacronym{nrdz}{NRDZ}{National Radio Dynamic Zone}
\newacronym{ofdm}{OFDM}{Orthogonal Frequency Division Multiplexing}
\newacronym{trl}{TRL}{technology readiness level}
\newacronym{ldgm}{LDGM}{low-density generator matrix}
\newacronym{ldpc}{LDPC}{low-density parity-check}
\newacronym{lo}{LO}{local oscillator}
\newacronym{isec}{ISEC}{Interdisciplinary Science and Engineering Complex}

\newacronym{osa}{OSA}{OpenAirInterface Software Alliance}
\newacronym{casper}{CASPER}{Collaboration for Astronomy Signal Processing and Electronics Research}
\newacronym{qos}{QoS}{Quality of Service}
\newacronym{oran}{O-RAN}{Open Radio Access Network}
\newacronym{ran}{RAN}{Radio Access Network}

\newacronym{ric}{RIC}{RAN Intelligent Controller}
\newacronym{cbrs}{CBRS}{Citizens Broadband Radio Service}
\newacronym{gaa}{GAA}{General Authorized Access}
\newacronym{pal}{PAL}{Priority Access Licensee}
\newacronym{fcc}{FCC}{Federal Communications Commission}
\newacronym{sas}{SAS}{spectrum access system}
\newacronym{ai}{AI}{artificial intelligence}
\newacronym{irs}{IRS}{intelligent reflecting surface}

\newacronym{ser}{SER}{symbol error rate}
\newacronym{rfic}{RFIC}{radio frequency integrated circuit}
\newacronym{rfi}{RFI}{\gls{rf} Interference}
\newacronym{aml}{AML}{adversarial machine learning}
\newacronym{sdn}{SDN}{software-defined networking}
\newacronym{star}{STAR}{Simultaneous Transmit and Receive}
\newacronym{sinr}{SINR}{signal-to-interference-noise ratio}
\newacronym{vlba}{VLBA}{Very Long Baseline Array}
\newacronym{ngvla}{ngVLA}{Next Generation Very Large Array}
\newacronym{nrao}{NRAO}{National Radio Astronomy Observatory}
\newacronym{fso}{FSO}{Free Space Optics}
\newacronym{ngso}{NGSO}{non-geostationary orbit}
\newacronym{vsd}{VSD}{Value-Sensitive Design}
\newacronym{sensr}{SENSR}{Spectrum Efficient National Surveillance Radar}
\newacronym{gbps}{Gbps}{Gigabit-per-second}
\newacronym{tbps}{Tbps}{Terabit-per-second}
\newacronym{nas}{NAS}{Network Attached Storage}
\newacronym{5gb}{5GB}{5G-and-beyond}
\newacronym{osi}{OSI}{Open Systems Interconnection}
\newacronym{onr}{ONR}{Office of Naval Research}
\newacronym{afosr}{AFOSR}{Air Force Office of Scientific Research}
\newacronym{afrl}{AFRL}{Air Force Research Laboratory}
\newacronym{arl}{ARL}{Army Research Laboratory}
\newacronym{bdss}{BDSS}{broadband directional spectrum sensor}
\newacronym{circ}{CIRC}{Community Infrastructure for Research in Computer and Information Science and Engineering}

\newacronym{aoa}{AoA}{Angle of Arrival}
\newacronym{noe}{NOE}{NRDZ Orchestration Engine}

\newacronym{mchem}{MCHEM}{Massive Channel Emulator}
\newacronym{afc}{AFC}{Automated Frequency Coordination}
\newacronym{esc}{ESC}{Environmental Sensing Capability}

\newacronym[firstplural=Devices Under Test (DUTs)]{dut}{DUT}{Device Under Test}

\newacronym{kpi}{KPI}{Key Performance Indicator}
\newacronym{dei}{DEI}{diversity, equity and inclusion}

\newacronym{itu}{ITU}{International Telecommunication Union}
\newacronym[firstplural=Notices of Inquiry (NOIs)]{noi}{NOI}{Notice of Inquiry}
\newacronym{wp}{WP}{Working Party}
\newacronym{drs}{DRS}{Digital Repository Service}
\newacronym{ms}{M.S.}{Master of Science}

\newacronym{lxc}{LXC}{Linux Container}
\newacronym{vpn}{VPN}{Virtual Private Network}
\newacronym{ldap}{LDAP}{Lightweight Directory Access Protocol}
\newacronym{ntn}{NTN}{non-terrestrial network}
\newacronym{xr}{XR}{eXtended Reality}
\newacronym{leo}{LEO}{low-Earth orbit}
\newacronym{hap}{HAP}{high-altitude platform}
\newacronym{isl}{ISL}{inter-satellite link}
\newacronym{ul}{UL}{uplink}
\newacronym{dl}{DL}{downlink}
\newacronym{vsat}{VSAT}{Very Small Aperture Terminal}
\newacronym{swap}{SWaP}{size, weight, and power}
\newacronym{cgr}{CGR}{Contact Graph Routing}
\newacronym{dtn}{DTN}{Delay Tolerant Network}
\newacronym{apd}{APD}{avalanche photo detector}
\newacronym{arq}{ARQ}{Automatic Repeat Request}
\newacronym{csli}{CSLI}{CubeSat Launch Initiative}
\newacronym{csdw}{CSDW}{CubeSat Developer Workshop}
\newacronym{ssc}{SSC}{Small Satellite Conference}
\newacronym{icc}{ICC}{International Conference on Communications}
\newacronym{api}{API}{Application Programming Interface}
\newacronym{3gpp}{3GPP}{3rd Generation Partnership Project}
\newacronym{ttc}{TT\&C}{Telemetry, Tracking and Control}
\newacronym{adcs}{ADCS}{attitude determination and control system}
\newacronym{pms}{PMS}{Power Management System}
\newacronym{ber}{BER}{Bit Error Rate}
\newacronym{cs}{CS}{Computer Science}
\newacronym{cise}{CISE}{Computer and Information Science and Engineering}
\newacronym{hpa}{HPA}{High Power Amplifier}
\newacronym{seu}{Single Event Upsets}{SEU}
\newacronym{ots}{OTS}{off-the-shelf}
\newacronym{msu}{MSU}{Morehead State University}
\newacronym{ppdr}{PPDR}{Public Protection and Disaster Relief}
\newacronym{sigmf}{SigMF}{Signal Metadata Format}
\newacronym{sscm}{SSC}{Space Science Center}
\newacronym{unp}{UNP}{University Nanosatellite Porgram}
\newacronym{unlab}{UNLab}{Ultrabroadband Nanonetworking Laboratory}
\newacronym{coe}{COE}{College of Engineering}
\newacronym{aeronu}{AeroNU}{AerospaceNU}
\newacronym{dsn}{DSN}{Deep Space Network}
\newacronym{click}{CLICK}{CubeSat Laser Infrared Cross-link}
\newacronym{tbird}{TBIRD}{Terabyte Infrared Delivery}
\newacronym{mc}{MC}{Mission Concept}
\newacronym{stem}{STEM}{Science, Technology Engineering, and Mathematics}
\newacronym{knrc}{KNRC}{Kostas Nanomanufacturing Research Center}
\newacronym{sdl}{SDL}{Space Dynamics Laboratory}
\newacronym{obc}{OBC}{On Board Computer}
\newacronym{gs}{GS}{Ground Station}
\newacronym{cots}{COTS}{commercial off-the-shelf}
\newacronym{ussf}{USSF}{U.S. Space Force}
\newacronym{usaf}{USAF}{U.S. Air Force}
\newacronym{iss}{ISS}{International Space Station}
\newacronym{bpsk}{BPSK}{Binary Phase Sfhit Keying}
\newacronym{conops}{ConOps}{Concept of Operations}
\newacronym{lsp}{LSP}{Launch Service Provider}
\newacronym{eps}{EPS}{Electric Power System}
\newacronym{cdh}{CD\&H}{Command and Data Handling}

\newacronym{scr}{SCR}{System Concept Review}
\newacronym{srr}{SRR}{System Requirement Review}
\newacronym{rvm}{RVM}{Requirement Verification Matrix}
\newacronym{pmr}{PMR}{Program Management Review}
\newacronym{pdr}{PDR}{Preliminary Design Review}
\newacronym{cdr}{CDR}{Critical Design Review}
\newacronym{fsr}{FSR}{Flight Selection Review}
\newacronym{pn}{PN}{Pseudorandom Noise}
\newacronym{mls}{MLS}{Maximum Length Sequence}
\newacronym{pdp}{PDP}{Power Delay Profile}
\newacronym{udp}{UDP}{User Datagram Protocol}
\newacronym{cfo}{CFO}{Carrier Frequency Offset}
\newacronym{eof}{EoF}{Enf of Frame}
\newacronym{sll}{SLL}{Side Lobe Level}
\newacronym{vhf}{VHF}{Very High Frequency}
\newacronym{uhf}{UHF}{Ultra High Frequency}
\newacronym{satnogs}{SATNOGS}{Satellite Networked Open Ground Station}
\newacronym{tle}{TLE}{Two-Line Element}
\newacronym{qam}{QAM}{Quadrature-Amplitude Modulation}
\newacronym{ssp}{SSP}{Satellite Service Provider}
\newacronym{raan}{RAAN}{Right Ascension of the Ascending Node}
\newacronym{cdf}{CDF}{Cumulative Density Function}
\newacronym{pdf}{PDF}{Probability Density Function}
\newacronym{lcm}{LCM}{Least Common Multiplier}

\newacronym{gds}{GDS}{Ground Data System}
\newacronym{geo}{GEO}{Geostationary}

\newacronym{tdrss}{TDRSS}{Tracking and Data Relay Satellite System}
\newacronym{nasa}{NASA}{National Aeronautics and Space Administration}
\newacronym{cnsa}{CNSA}{China National Space Administration}
\newacronym{marco}{MarCO}{Mars CubeSat One}
\newacronym{sso}{SSO}{Sun-synchronous orbit}
\newacronym{gsaas}{GSaaS}{Ground-Segment-as-a-Service}
\usepackage{balance}
\usepackage{cuted}
\usepackage{xcolor}
\usepackage{enumitem}
\usepackage{titlesec}

\titlespacing*{\subsection}
{0pt}                
{1.0ex plus .2ex minus .2ex} 
{0.5ex}      
\titlespacing*{\section}
{0pt}                
{1.5ex plus 1ex minus .2ex} 
{1.1ex plus .2ex}    

\usepackage{tikz}
\usepackage{pgfplots}
\usetikzlibrary{fit}
\makeatletter
\tikzset{
  fitting node/.style={
    inner sep=0pt,
    fill=none,
    draw=none,
    reset transform,
    fit={(\pgf@pathminx,\pgf@pathminy) (\pgf@pathmaxx,\pgf@pathmaxy)}
  },
  reset transform/.code={\pgftransformreset}
}
\makeatother

\begin{document}
\bstctlcite{IEEEexample:BSTcontrol}

\title{\vspace{-3mm}\LARGE Toward the Internet of Space Things:\,\,Performance Analysis of LEO~Satellite Relay Networks using mmWave and sub-THz links}

\author{\IEEEauthorblockN{Sergi Aliaga\textsuperscript{*},~\IEEEmembership{Student Member,~IEEE}, Ahmad Masihi\textsuperscript{*},~\IEEEmembership{Student Member,~IEEE}, Vitaly Petrov,~\IEEEmembership{Member,~IEEE}, Marc~Sanchez~Net,~\IEEEmembership{Member,~IEEE}, and Josep M. Jornet,~\IEEEmembership{Fellow,~IEEE}}
        \vspace{-11mm}
\thanks{S.~Aliaga, A.~Masihi, and J.~M.~Jornet are with Northeastern University, Boston, MA, USA. Email: \{aliaga.s,~masihi.a,~j.jornet\}@northeastern.edu
V.~Petrov is with KTH Royal Institute of Technology, Stockholm, Sweden. Email: vitalyp@kth.se. M.~S.~Net is with NASA Jet Propulsion Laboratory at California Institute of Technology, Pasadena, CA, USA. Email: marc.sanchez.net@jpl.nasa.gov. \textsuperscript{*}S.~Aliaga and A.~Masihi are co-first authors. The work by V.~Petrov has been supported by Vinnova 6GSTAR project. The work by M. S. Net was carried out at the Jet Propulsion Laboratory, California Institute of Technology, under a contract with the National Aeronautics and Space Administration (80NM0018D0004).
A~shorter version of this work has been presented at Asilomar~2024~\cite{aliaga2024non}.}}



\maketitle

\begin{abstract}
As the commercial space economy expands, existing ground-based infrastructure faces severe bottlenecks in supporting the data-intensive continuous connectivity needs of next-generation ``space users," including CubeSats, space data centers, and more. Even when utilizing existing Ku-band ground relay networks, the contact time with a CubeSat at low-Earth orbit (LEO) is often still limited to minutes per day only. This paper analyzes an alternative system design that leverages emerging high-rate millimeter-wave (mmWave) and sub-terahertz (sub-THz) inter-satellite links to build a high-throughput and high-availability satellite-based relay backbone for space vehicles. To evaluate this concept, we develop a comprehensive mathematical framework that jointly incorporates complex time-variant orbital dynamics and mmWave/sub-THz link characteristics. We then derive the key performance indicators, including contact probability, channel capacity, and energy efficiency. The numerical results, cross-verified by computer simulations, demonstrate that such systems can provide improvements of up to several orders of magnitude compared to existing networks of ground stations. Notably, we identify a fundamental bound on download capacity and show that continuous 24/7 connectivity becomes achievable with only ten LEO relay satellites. These findings establish mmWave and sub-THz satellite relay networks as a promising, scalable, and energy-efficient solution, thus unlocking improved connectivity with various space vehicles of tomorrow.
\vspace{-5mm}
\end{abstract}


\section{Introduction}\label{sec:introduction}
The satellite communications landscape has evolved from a few \gls{geo} satellites providing basic relay services into an ecosystem of massive \gls{leo} constellations comprising thousands of interconnected nodes. While legacy \gls{geo} systems suffered from high latency due to their 36,000~km altitude~\cite{quaglione1980evolution}, today's \gls{leo} mega-constellations operate between 400--2,000~km, slashing latency to under 10~ms and delivering fiber-like broadband performance~\cite{pachler2021updated}. Meanwhile, another ongoing trend in the industry to address continuously growing performance demands is a relatively slow but persistent adoption of new wider frequency bands for the satellite-to-ground and inter-satellite links.

While selected millimeter-wave bands (mmWave, such as Ka-band and Ku-band~\cite{pachler2021updated}) are already widely used in state-of-the-art systems~\cite{kodheli2021satellite}, other candidate bands are also receiving growing attention from academia and industry. These include but are not limited to sub-terahertz (sub-THz, 100~GHz--300~GHz)~\cite{De_Gaudenzi2022Open}, terahertz (THz, 300~GHz--3~THz)~\cite{meltem2021terahertz,aliaga2022joint}, and free-space optics (FSO,~190~THz--230~THz)~\cite{chaudhry2022temporary}. These emerging \gls{leo} constellations utilizing wideband wireless links are expected to become an essential component of forthcoming \glspl{ntn} -- integrated systems of satellites, airborne, and terrestrial nodes -- which serve as a cornerstone for 6G and beyond by promising ubiquitous connectivity across land, sea, and air.

\subsection{A vision of better connectivity for space users}
Today, the overwhelming majority of research efforts on \glspl{ntn} is focused on providing better services for \emph{terrestrial} nodes (e.g., smartphones and cars), \emph{marine} traffic (various types of vessels), and \emph{airborne} users (e.g., planes and UAVs)~\cite{Azari_survey_2022}. However, there exists another group of \gls{ntn} beneficiaries -- currently small in volume but extensively growing -- \emph{emerging ``space users''}, which are the main focus of this work. These users primarily represent various \gls{leo} satellites owned by companies, research centers, and government bodies, but can ultimately also include orbital data centers, lunar landers, and space telescopes, among others~\cite{Akyildiz_iost_2019}.

Such a population of space users is expanding rapidly; as of 2026, there are approximately 15,000 active payloads in orbit, a nearly fourfold increase from just five years ago~\cite{ESA_SpaceEnv_2025}. Further, modern space vehicles can carry advanced imaging payloads generating terabytes of data per mission~\cite{tiede2018big}, yet current \glspl{gds} cannot accommodate this exponential demand. Even NASA's \gls{dsn} faces projected capacity shortfalls as client requirements escalate~\cite{abraham2019recommendations}. This bottleneck stems from both spectrum exhaustion in the Ku-band and ground station scheduling conflicts, with access wait times reaching hours or days.

\begin{figure*}[t]
    \centering
    \vspace{-5mm}
    \includegraphics[width=0.99\textwidth]{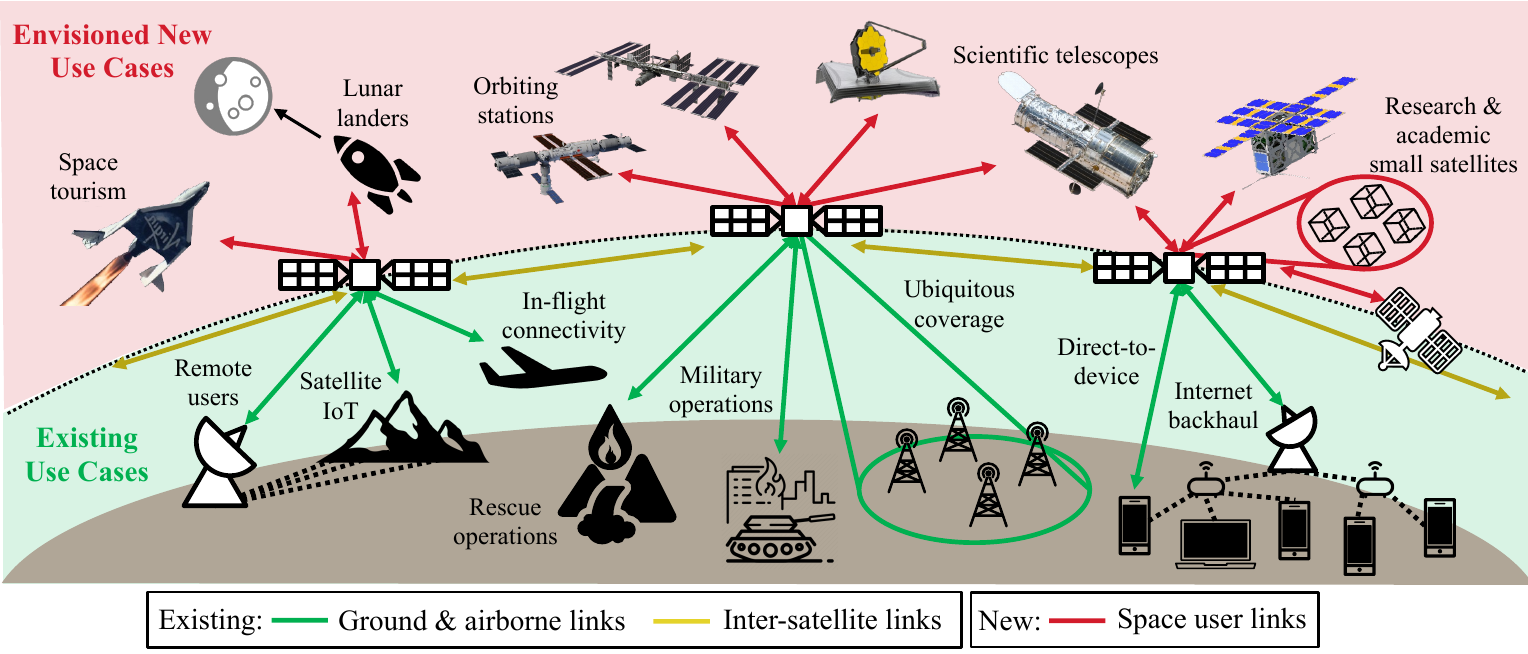}
    \vspace{-2mm}
    \caption{Envisioned scope of 6G+ \gls{ntn} connectivity to support \emph{space users} (scientific telescopes and missions, space tourism, lunar landers, and orbital stations, etc.), alongside traditional \emph{terrestrial and airborne users}.}
    \label{fig:vision}
    \vspace{-4mm}
\end{figure*}

We aim to address this limitation in the present study by exploring the following question: \emph{Why not leverage existing and future high-rate \gls{leo} satellite communication networks to also serve space-based users?} First, such advanced networks, illustrated in Fig.~\ref{fig:vision}, can relay data to the backbone with minimal disturbance, ensuring timely delivery to mission centers and scientific teams. Aligning with 3GPP standards for NTN-ground integration~\cite{3GPP2020Study}, adopting standardized SIM-equivalent radio interfaces would simplify space user hardware, thereby reducing costs and weight and potentially improving scalability, capacity, and energy efficiency.

Beyond technical gains, this \gls{ntn}-centric approach leverages existing commercial infrastructure investment, allowing constellations deployed for terrestrial markets to serve space users without dedicated relay satellites. The resulting space-to-\gls{ntn} traffic imposes minimal incremental load compared to terrestrial demand, facilitating efficient spectrum sharing—especially over oceans or unpopulated regions where ground user density is low. Finally, properly configured constellations provide continuous coverage, eliminating the scheduling conflicts and communication gaps inherent in sparse ground station networks.

\subsection{Related work}
In principle, the idea of using satellites for space data relaying is not entirely new and has been explored before through several successful infrastructures. For instance, NASA's \gls{tdrss} has utilized \gls{geo} satellites for over 30 years to provide Ku-band coverage, with ongoing plans to integrate optical communications~\cite{israel2018next}. Similarly, the \gls{cnsa} employs lunar relay satellites in the S- and Ku-bands~\cite{zhang2021development}, while the \gls{marco} mission recently demonstrated the first use of CubeSats to relay scientific data from Mars during critical mission phases~\cite{Schoolcraft2017Marco}.

These early successes have motivated numerous studies on relay architectures for near-Earth and deep-space missions. Wan et al.~\cite{wan2019structured} modeled small solar system relay constellations, while Modenini et al.~\cite{modenini2022two} proposed relays at Lagrange points and Cheung et al.~\cite{cheung2023deep} studied heliocentric orbits  to mitigate deep-space ground infrastructure and optical conjunction challenges. Elewaily et al.~\cite{elewaily2024delay} surveyed \gls{dtn} frameworks as a solution for link disruptions, though these are typically limited to $<100$ nodes and have not been evaluated for massive next-generation \glspl{ntn}. Most of these works focus on dedicated relay constellations operating below 100~GHz and orbiting specific planetary bodies. Palermo et al.~\cite{palermo2015earth} considered an Earth-orbiting relay system for space users, but it relied on dedicated architectures operating in the Ku-band.

Notably, numerous works explore beyond-100~GHz technology in next-generation \glspl{ntn}~\cite{mehdi2018thz}, but they predominantly focus on ground or airborne users. Al-qaraghuli et al.~\cite{alqaraghuli2021performance} proposed a dual Sub-THz/Ka-band system to provide high-rate Internet to ground users via massive \gls{leo} constellations, while Nie et al.~\cite{nie2021channel}  characterized THz inter-satellite links for ultra-high throughput between network nodes. For airborne coverage, literature includes detailed THz channel models~\cite{kokkoniemi2021channel}, stochastic geometry frameworks for satellite-to-airplane downlinks~\cite{wang2023coverage}, and broad reviews of THz aerospace opportunities~\cite{gao2025terahertz}. However, these studies ignore the specific peculiarities (e.g., mobility and orbital dynamics) of the growing space users segment. 

\subsection{Claims and contributions}
Despite the intuitive advantages of serving space users through (sub-)THz-enabled \glspl{ntn} (\cite{hwu2013terahertz}, among others), \emph{a comprehensive methodology to quantify the performance boundaries for space users served by mmWave/sub-THz-capable \glspl{ntn} remains missing}. Hence, by extending our preliminary work in~\cite{aliaga2024non}, we aim to address this gap. The main contributions of this study are thus summarized as follows:
\begin{itemize}
    \item \emph{Flexible Mathematical Framework}: An elaborate framework is developed to quantify \gls{ntn}-based space user capacity. The developed framework takes into account all the major static and time-variant orbital parameters, the satellites' mutual orientation and mutual mobility, as well as the essential radio link parameters.
    \item \emph{In-depth Numerical Analysis}: A thorough investigation is performed using the developed methodology of the benefits of utilizing existing \gls{ntn} architectures to serve space users with state-of-the-art mmWave and sub-THz technology. The study highlights time-variant, statistical, and time-averaged \glspl{kpi}, including the probability of contact, total download capacity, and energy efficiency. The study particularly emphasizes the deployment configurations and orbital parameters that maximize the outlined \glspl{kpi}.
    \item \emph{Comprehensive Simulation Study}: The results delivered with the mathematical framework are further cross-verified by system-level computer simulation. We utilize current Ku-band ground data systems as a baseline to reveal substantial improvements in the outlined \glspl{kpi} with the considered concept.
\end{itemize}

The remainder of this paper details the system model and KPIs in ~\ref{sec:system_model}, derives the mathematical framework in Sec.~\ref{sec:analysis}, and presents the numerical results and conclusions in Sec.~\ref{sec:results} and ~\ref{sec:conclusions}, respectively

\section{System Model}
\label{sec:system_model}
\vspace{-1mm}
In this section we present the foundational assumptions underlying our model, encompassing the considered system architecture, the adopted propagation and network models, and the key performance metrics of interest. All variables and parameters introduced throughout this work are summarized in Table~\ref{tab:variables} for reference.
\begin{table}[!b]
\vspace{-2mm}
\caption{Summary of the notation used.}
\vspace{-1mm}
\centering
\begin{tabular}{p{0.1\textwidth}p{0.32\textwidth}}
\hline
\textbf{Variable} & \textbf{Description} \\
\hline
&\textbf{\textit{Physical constants}} \\
\hline
$\varsigma$ & Speed of light \\
$k$ & Boltzmann’s constant \\
$R_{\text{E}}$, $M_{\text{E}}$ & Earth radius and mass \\
$G$ & Gravitational constant\\
$\mu$ & Earth's standard gravitational parameter\\
\hline
&\textbf{\textit{Radio and propagation}} \\
\hline
$P_{\mathrm{Tx}}, P_{\mathrm{Rx}}$, $P_{\mathrm{N}}$  & Transmitted, received, and noise power \\
$G_{\mathrm{CS}}, G_{\mathrm{Rx}}$ & CubeSat and Receiver antenna gain \\
$f_c$ & Carrier frequency \\
$W$ & Bandwidth \\
$d(t)$ & Distance between CubeSat and receiver\\
$L, L_{\mathrm{spr}}, L_{\mathrm{abs}}$ & Path loss, spreading loss and absorption loss \\
$Q, p$ & Atmospheric composition and pressure \\
$T_0, T_{\mathrm{atm}} $ & Reference and atmospheric temperature\\
$\kappa$ & Molecular absorption coefficient \\
$r$ & propagation path range\\
$T_{\mathrm{a}}, T_{\mathrm{sys}}$ & Antenna and total system noise temperature\\
$F_{\mathrm{N}}$ & Receiver noise figure \\
$\rho$ & Signal-to-noise ratio \\
\hline
&\textbf{\textit{System model}} \\
\hline
$T_{\mathrm{c}}$ & Contact time between CubeSat and receiver \\
$T_{\mathrm{day}}, T$ & Earth day and scenario repetition period \\
$T_{\mathrm{res}}$ & Temporal resolution used for the calculation of $T$ \\
$N_{\text{S}}$, $N_{\text{G}}$ & Number of NTN relays and GSs\\
$N_{\text{S}}^*$ & Min. number of NTN relays for no outage\\
$\omega$ & Angular velocity\\
$R$, $h$ & Orbital radius and altitude ($R=R_{\mathrm{e}}+h$)\\
$d^{\mathrm{min}}$, $d^{\mathrm{max}}$ & Minimum and maximum possible link distances\\
$d^{\mathrm{B}}$ & Link blockage distance (blockage by the Earth)\\
$V$ & Visibility indicator function\\
$f, F$ & Probability and cumulative density functions \\
$i$ & Orbital inclination \\
$\Omega$ & Right Ascension of the Ascending Node (RAAN) \\
$M$ & Conversion matrix between 2-D and 3-D systems\\
$x, y, z$ & Orbital coordinates in the 3-D coordinate system\\
$\varphi, \Lambda$ & Ground station latitude and longitude\\
$\beta$ & Rotation of the Earth w.r.t. 3D coordinate system \\
$\theta$, $\theta_{\mathrm{th}}$ & GS elevation angle and Min. GS elevation angle\\
$\tau$, $\xi$ & Simple roots of $g(\hspace{-0.2mm}t\hspace{-0.2mm})\hspace{-1mm}=\hspace{-1mm}d_{\mathrm{GS}}(\hspace{-0.2mm}t\hspace{-0.2mm})-d_{\mathrm{GS}}^{B}$ and $d_{\mathrm{GS}}(\hspace{-0.2mm}t\hspace{-0.2mm})-d$ \\
$u(t)$ & Unit-step function \\
\hline
&\textbf{\textit{Performance metrics}} \\
\hline
$Q$ & Contact probability\\
$C$ & Channel capacity\\ 
$\Gamma$ & Download capacity over 24h\\
$\eta$ & Energy efficiency \\
\hline
\end{tabular}
\label{tab:variables}
\end{table}

\subsection{Deployment scenarios}
\begin{figure}[t]
    \centering
    \includegraphics[width=\linewidth]{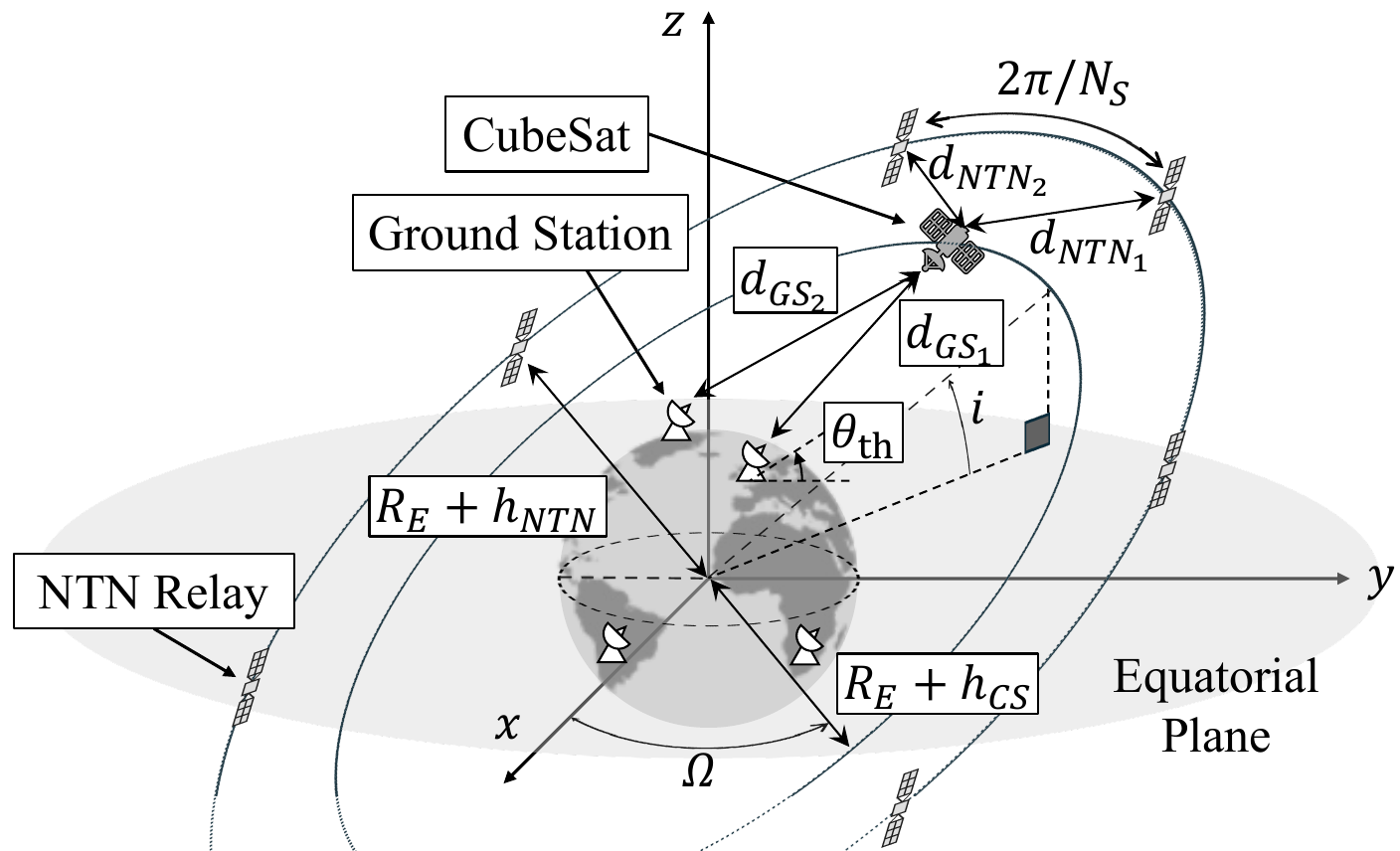}
    \caption{Modeling the coverage of a CubeSat (space user) using NTNs, compared to GS-based service.}
    \vspace{-6mm}
    \label{fig:sysmodel}
\end{figure}

The considered deployment scenarios, depicted in Fig.~\ref{fig:sysmodel}, involve a \gls{leo} CubeSat at an altitude $h_{\mathrm{CS}} \in [400, 2000]$~km above Earth ($R_E$). The CubeSat requires a reliable high-rate downlink for scientific data, including measurements, imagery, and telemetry. We evaluate \emph{four} scenarios to characterize downlink capacity, temporal dynamics, and total data volume, comparing traditional GS-based baselines with emerging NTN mega-constellations. These architectures include:

\begin{enumerate}
    \item \textbf{Single Ground Station:} The CubeSat communicates directly with a single terrestrial \gls{gs} ($N_{\text{G}}=1$). This represents the baseline architecture, where connectivity is limited to the visibility windows of one fixed \gls{gs}.
    
    \item \textbf{Multiple Ground Stations:} The CubeSat gets served by a ground relay network of $N_{\text{G}} > 1$ GSs. This scenario represents another typical approach, where a mission relies on a network of geographically distributed \glspl{gs} (typically, third-party) to improve coverage and latency at the cost of increased operational expenses.
    
    \item \textbf{Single NTN Relay:} The CubeSat sends its data through a single \gls{ntn} satellite relay ($N_{\text{S}} = 1$), which then communicates with the \gls{ssp} ground infrastructure. This configuration facilitates direct comparison of one \gls{gs} (Scenario~1) vs. one \gls{ntn} relay.
    
    \item \textbf{Multiple NTN Relays:} This architecture generalizes the previous single-relay scenario by employing $N_{\text{S}}$ \gls{leo} relays in the same orbit. Similar to multi-GS Scenario~2, this setup is expected to provide increased spatial and temporal diversity, more frequent access opportunities, and enhanced reliability.
\end{enumerate}

For clarity and tractability of our first-order analysis, we assume that the \gls{ntn} relays' orbit is co-planar to the CubeSat orbit. The rationale here is that typical satellite constellations following a Walker-Delta configuration are composed of multiple orbital planes to provide uniform global coverage~\cite{walker1984satellite}. Due to the high density of these planes, a CubeSat's orbit remains approximately co-planar with that of the nearest relay satellites most of the time, even with orbital precession. This justifies the selected co-planar deployment as a tractable yet representative approximation of the real system dynamics.

As illustrated in Fig.~\ref{fig:sysmodel}, we adopt an Earth-centered, right-handed Cartesian coordinate system where the $z$-axis aligns with Earth's rotational axis and the $x$-axis points toward the vernal equinox. The CubeSat and \gls{ntn} relays occupy circular LEO orbits in the same plane at altitudes $h_{\mathrm{CS}}$ and $h_{\mathrm{NTN}}$, with Earth-center distances $R_{\mathrm{E}} + h_{\mathrm{CS}}$ and $R_{\mathrm{E}} + h_{\mathrm{NTN}}$, respectively. Since feeder links are assumed to be seamlessly integrated with terrestrial infrastructure, their detailed modeling is omitted. While Fig.~\ref{fig:sysmodel} depicts the CubeSat at a lower altitude for clarity, the model applies to any orbital configuration. Relative positions are defined by orbital inclination $i$ and \gls{raan} $\Omega$. Time-varying communication distances $d_{\mathrm{GS}}$ and $d_{\mathrm{NTN}}$ drive the scenario dynamics as satellites move along their orbits. 

\vspace{-2mm}
\subsection{Propagation and routing assumptions}
We model the wireless signal propagation using the Friis transmission equation, which describes the received power at a \gls{gs} or \gls{ntn} relay as:
\vspace{-1mm}
\begin{equation}\label{eq:rxPower}
    P_{\mathrm{Rx}} = \frac{P_{\mathrm{Tx}} G_{\mathrm{CS}} G_{\mathrm{Rx}}}{L(f_c, d(t))},
\end{equation}

where $P_{\mathrm{Tx}}$ is the CubeSat's transmit power, $G_{\mathrm{CS}}$ is its antenna gain, and $G_{\mathrm{Rx}}$ denotes the gain of the receiving node (either a GS or \gls{ntn} relay). Mutual antenna alignment between the transmitter and receiver is assumed. The channel path loss $L(f_c, d(t))$ depends on the carrier frequency $f_c$ and the time-varying distance $d(t)$ between the CubeSat and the receiver.

For the CubeSat-to-\gls{ntn} relay link, the propagation occurs entirely in space. As a result, no atmospheric absorption is encountered, and the only relevant attenuation mechanism is free-space spreading loss, modeled as:
\vspace{-2mm}
\begin{equation}
    L_{\mathrm{spr}}(f_c, d(t)) = \left( \frac{4 \pi d(t) f_c}{\varsigma} \right)^2,
\end{equation}
where $\varsigma$ is the speed of light.

In contrast, the CubeSat-to-GS link traverses the atmosphere, where molecular absorption may occur. We model the corresponding molecular absorption loss as~\cite{ITU676}:
\vspace{-2mm}
\begin{equation}
    L_{\mathrm{abs}}(f_c, d(t)) = \exp \left[ \int_0^{d(t)} \hspace{-5mm}\kappa(f_c, Q(r), p(r), T_{\mathrm{atm}}(r)) \, dr \right]\hspace{-1mm},
\end{equation}
where $\kappa$ is the frequency-dependent molecular absorption coefficient, influenced by the atmospheric composition $Q(r)$, pressure $p(r)$, and temperature $T(r)$ along the path $r$. These atmospheric quantities are obtained from ITU-R Recommendation P.835~\cite{ITU835}. Therefore, the total path loss for the CubeSat-to-GS link is modeled as:
\vspace{-1mm}
\begin{equation}
    L(f_c, d(t)) = L_{\mathrm{spr}}(f_c, d(t)) L_{\mathrm{abs}}(f_c, d(t)).
\end{equation}
\vspace{-4mm}

To analyze the capacity of relaying data from the CubeSat to a multi-relay \gls{ntn} system, we assume that the CubeSat always transmits to the nearest visible relay whenever possible. A similar assumption is made for CubeSat-to-GS links, where the CubeSat always communicates with the nearest \gls{gs} in line of sight.

\subsection{Metrics of interest}
\label{subsec:metrics}
In this study, we evaluate upper bounds for system-level performance, accounting for both wireless technology and architectural alternatives. We primarily focus on the following \glspl{kpi}:

\paragraph*{a) Contact probability, $Q$} Defined as the probability that the CubeSat is in \gls{los} with either a \gls{gs} or an \gls{ntn} relay at a given time. It is calculated as the ratio between the total contact duration between CubeSat and the GS or \gls{ntn} relay, $T_c$, and the scenario repetition period, $T$:
\vspace{-1mm}
\begin{equation}
\label{eq:q}
Q = \frac{T_c}{T}.
\end{equation}
\vspace{-2mm}

A further detailed discussion and formulation for $T_c$ and $T$ is provided in Sec.~\ref{sec:analysis}.

\paragraph*{b) Instantaneous channel capacity, $C(t)$} This time-varying metric is derived using Shannon’s capacity formula:
\vspace{-1mm}
\begin{equation}\label{eq:capacity_general}
C(t) = W \log_2 \left( 1 + \rho(t) \right) = W \log_2 \left( 1 + \frac{P_{\mathrm{Rx}}(t)}{P_{\mathrm{N}}} \right),
\end{equation}
\vspace{-1mm}
where $W$ is the system bandwidth, $\rho(t)$ is the instantaneous \gls{snr}, $P_{\mathrm{Rx}}(t)$ is the received power at time $t$, and $P_{\mathrm{N}}$ is the system noise power. The received power varies primarily due to changes in the distance $d(t)$ caused by the CubeSat's orbital motion.

The receiver noise power is calculated as $P_{\mathrm{N}} = kT_{\text{sys}}W$, where $k$ is the Boltzmann constant, and $T_{\text{sys}}$ is the system noise temperature. Accordingly, the total system noise temperature $T_{\text{sys}}$ is expressed as:
\vspace{-1mm}
\begin{equation}
    T_{\text{sys}} =  \left( 10^{F_{\mathrm{N}}/10} - 1 \right)T_0 + T_a,
\end{equation}
\vspace{-1mm}
where $T_0$ is the reference temperature (typically 290~K), $F_{\mathrm{N}}$ is the receiver noise figure in dB, and $T_a$ is the antenna noise temperature, which depends on the receiver's field of view and pointing direction. 

\paragraph*{c) Download capacity, $\Gamma$} This metric represents the total amount of data that can be downloaded over a 24-hour period. It is computed by averaging the channel capacity over the scenario repetition period and scaling accordingly:
\vspace{-1mm}
\begin{equation}
\label{downloadcap}
\Gamma = \frac{T_{\text{day}}}{T} \int_0^T C(t) \, dt.
\end{equation}
\vspace{-1mm}

\paragraph*{d) Energy efficiency, $\eta$} Defined as the ratio of the total useful data transmitted to the energy expended. Assuming the CubeSat transmits with a constant power $P_{\mathrm{Tx}}$ whenever a link is available, the energy efficiency is:
\vspace{-1mm}
\begin{equation}
\label{eq:energyeff_general}
\eta = \frac{1}{P_{\mathrm{Tx}} T_c} \int_0^T C(t) \, dt.
\end{equation}
\vspace{-2mm}

This metric reflects the system's effectiveness in delivering data relative to its energy consumption, which is particularly critical for power- and battery-constrained space users.

\section{Mathematical Framework}
\label{sec:analysis}
In this section, we present the mathematical analysis of the performance metrics for the system model considered. 

\subsection{Distance to CubeSat}\label{subsec:dist2cubesat}
We first analyze the time-varying geometric distance between the CubeSat and the two receiver types—\glspl{gs} and \gls{ntn} relays. This analysis is fundamental as it directly influences all performance metrics defined in Sec.~\ref{sec:system_model}

\subsubsection{CubeSat-to-Ground Station Distance}
The distance $d_{\mathrm{GS}}(t)$ varies due to the CubeSat's orbital motion and the Earth's rotation. To compute $d_{\mathrm{GS}}(t)$, we transform the CubeSat's position from its 2-D orbital plane to the 3-D Earth-centered coordinate system. This involves rotating by the \gls{raan} $\Omega$ around the $z$-axis and the inclination $i$ around the $x$-axis. The resulting transformation matrix is: 
\begin{equation}
\label{eq:matrix_transform}
 M=
\begin{bmatrix}
\cos \Omega & \sin \Omega & 0 \\
-\cos i \sin \Omega & \cos i \cos \Omega & \sin i \\
\sin i \sin \Omega & -\sin i \cos \Omega & \cos i
\end{bmatrix}
\end{equation}

The CubeSat’s position in the 3-D coordinate system is then computed using its position in the 2-D coordinate system as:
\begin{equation}
\label{eq:inv_matrix_transform}
\begin{bmatrix}
x_{\text{CS}} \\
y_{\text{CS}} \\
z_{\text{CS}}
\end{bmatrix} = 
M^{-1}
\begin{bmatrix}
R_{\text{CS}} \cos(\omega_{\text{CS}}t) \\
R_{\text{CS}} \sin(\omega_{\text{CS}}t) \\
0
\end{bmatrix}.
\end{equation}

On the other hand, the position of the \gls{gs} is defined in the 3-D coordinate system by:
\vspace{-1mm}
\begin{equation}
\label{eq:GS_coordinates}
\begin{cases}
x_{\text{GS}} = R_{\text{E}} \cos \varphi \cos(\Lambda + \omega_{\text{GS}}t - \beta ) \\
y_{\text{GS}} = R_{\text{E}} \cos \varphi \sin(\Lambda + \omega_{\text{GS}}t - \beta) \\
z_{\text{GS}}= R_{\text{E}} \sin \varphi
\end{cases},
\end{equation}
where $\varphi$ and $\Lambda$ represent the latitude and longitude of the \gls{gs}, respectively, and the parameter $\beta$ represents the angular rotation of the Earth with respect to the 3-D coordinate system at the adopted time origin. The resulting Euclidean distance is then calculated as:
\begin{equation}
\label{eq:dgs}
d_{\text{GS}}(t) = \sqrt{(x_{\text{GS}} \hspace{-1mm}- \hspace{-1mm}x_{\text{CS}})^2 + (y_{\text{GS}}\hspace{-1mm} -\hspace{-1mm} y_{\text{CS}})^2 + (z_{\text{GS}}\hspace{-1mm} - \hspace{-1mm}z_{\text{CS}})^2},
\end{equation}
where the dependency of the position components with time has been committed for clarity.

Based on this, the minimum and maximum possible distances are defined by:
\begin{equation}
\label{eq:dgsmin}
\hspace{-2mm}d_{\text{GS}}^{\min}\hspace{-1mm}= \hspace{-1mm}
\begin{cases}
\sqrt{R_{\text{E}}^2 + R_{\text{CS}}^2 - 2R_{\text{E}}R_{\text{CS}} \cos(|\varphi| - i)}, &\hspace{-1mm} |\varphi| \geq i \\
h_{\text{CS}}, &\hspace{-1mm} |\varphi| \leq i
\end{cases}
\end{equation}
\vspace{-1.5mm}
\begin{equation}
\label{eq:dgsmax}
\hspace{-2mm}d_{\text{GS}}^{\max}\hspace{-1mm}= \hspace{-1mm}
\begin{cases}
\sqrt{R_{\text{E}}^2 + R_{\text{CS}}^2 + 2R_{\text{E}}R_{\text{CS}} \cos(|\varphi| - i)}, & \hspace{-1mm}|\varphi| \geq i \\
2R_{\text{E}}+h_{\text{CS}}, &\hspace{-1mm} |\varphi| \leq i
\end{cases}.
\end{equation}

Notably, we analyze $d_{\mathrm{GS}}(t)$ regardless of the Earth's blockage, which we then take into account as a mask to the possible values of $d_{\mathrm{GS}}(t)$. In this regard, we define the corresponding maximum blockage distance between the CubeSat and the \gls{gs}:
\vspace{-2mm}
\begin{equation}
\label{eq:dgsv}
\begin{split}
d_{\text{GS}}^{\text{B}} = &\biggr[R_{\text{E}}^2 + R_{\text{CS}}^2 - \\
&2R_{\text{CS}}R_{\text{E}} \sin\left(\theta_{\mathrm{th}} + \sin^{-1}\left(\frac{R_{\text{E}}}{R_{\text{CS}}} \cos \theta_{\mathrm{th}}\right)\right)\biggr]^{1/2},
\end{split}
\end{equation}
where $\theta_{\mathrm{th}}$ indicates the minimum elevation angle of a \gls{gs} given the surrounding obstacles. Because we are restricting the present analysis to just circular \gls{leo} orbits, it is always the case that $d_{\text{GS}}^{\text{B}}<d_{\text{GS}}^{\max}$. Using the notions of $d^{\min}_{\text{GS}}$, and $d^{\text{B}}_{\text{GS}}$, we derive the following visibility function:
\begin{equation}
\label{eq:vdgs}
V_{\text{GS}}(d_{\text{GS}}) = 
\begin{cases}
1, & d_{\text{GS}}^{\min} < d_{\text{GS}} < d_{\text{GS}}^{\text{B}} \\
0, & \text{Otherwise}.
\end{cases}
\end{equation}
\vspace{-5mm}

Since no closed-form solution exists for the visibility inequality when incorporating \eqref{eq:dgs} into \eqref{eq:vdgs}, visibility periods must be computed numerically. Let $\tau_1, \tau_2, \ldots, \tau_L$ be the simple roots of $g(t) = d_{\mathrm{GS}}(t) - d_{\mathrm{GS}}^\text{B}$ within $[0,\, T_{\mathrm{GS}}]$, such that $g(\tau_l) = 0$ and $g'(\tau_l) \neq 0$. The resulting visibility function $V_{\mathrm{GS}}(t)$ is defined in \eqref{eq:vtgs}, where $u(t)$ is the unit step function and $g'(t)$ is the derivative of $g(t)$.

Using $d_{\text{GS}}^{\max}$, we define the \gls{cdf} of the distance from the CubeSat to the \gls{gs}, assuming that $t\sim U[0, T_{\text{GS}}]$:
\vspace{-1mm}
\begin{equation}
\label{eq:Fdgs}
F_{d_{\text{GS}}}(d) = 
\begin{cases}
0, & d \leq d_{\text{GS}}^{\min} \\
F_{d_{\text{GS}}}^*(d), & d_{\text{GS}}^{\min} < d < d_{\text{GS}}^{\max} \\
1, & d \geq d_{\text{GS}}^{\max},
\end{cases}
\end{equation}
where $F_{d_{\mathrm{GS}}}^*$ is defined as follows:
\begin{equation}
\label{eq:FstardGs}
F_{d_{\mathrm{GS}}}^*(d) = P \left( d_{\text{GS}}(t) \leq d\right) = P \left( d_{\text{GS}}(t)-d \leq 0\right).
\end{equation}

Calculating $F_{d_{\mathrm{GS}}}^*(d)$ requires identifying the simple roots $\xi_1, \xi_2, \ldots, \xi_{K}$ of the inequality in~\eqref{eq:FstardGs}. As shown in~\eqref{eq:Fstar2}, the expression is derived by integrating over all time intervals within $T_{\mathrm{GS}}$ where $d_{\text{GS}}(t) \le d$.

If multiple \glspl{gs} are considered, the CubeSat connects to the closest visible station. Thus, $d_{\mathrm{GS}}(t)$ is defined as the minimum distance between the CubeSat and the $N_{\text{G}}$ \glspl{gs}:
\begin{equation}
\label{eq:dmgs}
d_{\text{GS}}(t) = \min\{d_{\text{GS}_1}(t), d_{\text{GS}_2}(t),...,d_{\text{GS}_{N_{\text{G}}}}(t)\},
\end{equation}
and the rest of the analysis is equally derived. 

\subsubsection{CubeSat-to-\gls{ntn} Relays Distance}
\label{subsubsec:ntn_distance}
$d_{\mathrm{NTN}}(t)$ is a periodic function of the relative motion between co-planar orbits. Assuming alignment at $t = 0$ and $N_{\text{S}}$ evenly spaced relays, the angular distance between consecutive nodes is $2\pi/N_{\text{S}}$. As the CubeSat overtakes (or is overtaken by) the relays due to altitude-dependent angular velocities, the link distance repeats with period $T_{\text{NTN}}$. This period, representing the time to sweep a relative angular displacement of $2\pi/N_{\text{S}}$, depends on $N_{\text{S}}$ and the relative angular velocity $\omega_{\mathrm{CS}} - \omega_{\mathrm{NTN}}$:
\vspace{-1mm}
\begin{equation}
\label{eq:Tdntn}
T_{\text{NTN}} = \frac{2\pi}{N_{\text{S}} \left| \omega_{\text{CS}} - \omega_{\text{NTN}}\right|}.
\end{equation}
The angular velocities of the CubeSat and the \gls{ntn} relays are derived from Kepler's third law, given by: 
\begin{equation}
\label{eq:angular_velocities}
\omega_{\text{CS/NTN}} = \sqrt{\frac{\mu}{R_{\mathrm{\text{CS/NTN}}}^3}}, 
\end{equation}
\vspace{-7mm}
\begin{strip}
\rule{\textwidth}{0.4pt} 
\begin{equation}
\label{eq:vtgs}
V_{\mathrm{GS}}(t) = 
\begin{cases}
\sum\limits_{n=1}^{L/2} u(t - \tau_{2n-1}) - \sum\limits_{n=1}^{L/2} u(t - \tau_{2n}),  
& \text{if } g'(\tau_{1}) < 0 \text{ and } g'(\tau_L) > 0 \\[1ex]

u(t) + \sum\limits_{n=1}^{(L-1)/2} u(t - \tau_{2n}) - \sum\limits_{n=1}^{(L+1)/2} u(t - \tau_{2n-1}),  
& \text{if } g'(\tau_{1}) > 0 \text{ and } g'(\tau_L) > 0 \\[1ex]

\sum\limits_{n=1}^{(L+1)/2} u(t - \tau_{2n-1}) - \sum\limits_{n=1}^{(L-1)/2} u(t - \tau_{2n}) - u(t - T_{\mathrm{GS}}),  
& \text{if } g'(\tau_{1}) < 0 \text{ and } g'(\tau_L) < 0 \\[1ex]

u(t) + \sum\limits_{n=1}^{L/2} u(t - \tau_{2n}) -\sum\limits_{n=1}^{L/2} u(t - \tau_{2n-1}) 
- u(t - T_{\mathrm{GS}}),  
& \text{if } g'(\tau_{1}) > 0 \text{ and } g'(\tau_L) < 0 \\[1ex]
\end{cases}
\end{equation}
\end{strip}
where $\mu = G M_{\text{E}}$ is the Earth's standard gravitational parameter, $G$ is the gravitational constant and $M_{\text{E}}$ the mass of Earth ($\mu \approx 3.986 \times 10^{14} \ \mathrm{m^3/s^2}$).
\begin{figure*}[!b]
\centering
\vspace{-5mm}
\rule{\textwidth}{0.4pt} 
\begin{minipage}{0.9\textwidth}
\vspace{2mm}
\begin{equation}
\label{eq:Fstar2}
F_{d_{\mathrm{GS}}}^*(d) =
\begin{cases}
\frac{1}{T_{\mathrm{GS}}}\left(\sum\limits_{n=1}^{\frac{K}{2}}\xi_{2n} -
\sum\limits_{n=1}^{\frac{K}{2}} \xi_{2n-1}\right), 
& \text{if } g'(\xi_{1}) < 0 \text{ and } g'(\xi_{K}) > 0 \\[1ex]

\frac{1}{T_{\mathrm{GS}}}\left(\sum\limits_{n=1}^{\frac{K+1}{2}} \xi_{2n-1} -
\sum\limits_{n=1}^{\frac{K-1}{2}} \xi_{2n}\right), 
& \text{if } g'(\xi_{1}) > 0 \text{ and } g'(\xi_{K}) > 0 \\[1ex]

\frac{1}{T_{\mathrm{GS}}}\left(T_{\mathrm{GS}} +
\sum\limits_{n=1}^{\frac{K-1}{2}} \xi_{2n} -
\sum\limits_{n=1}^{\frac{K+1}{2}} \xi_{2n-1}\right), 
& \text{if } g'(\xi_{1}) < 0 \text{ and } g'(\xi_{K}) < 0 \\[1ex]

\frac{1}{T_{\mathrm{GS}}}\left(T_{\mathrm{GS}} +
\sum\limits_{n=1}^{\frac{K}{2}} \xi_{2n-1} -
\sum\limits_{n=1}^{\frac{K}{2}} \xi_{2n}\right), 
& \text{if } g'(\xi_{1}) > 0 \text{ and } g'(\xi_{K}) < 0 \\[1ex]

\end{cases}
\end{equation}
\end{minipage}
\end{figure*}
The distances between the CubeSat and two consecutive relays are calculated via the law of cosines in \eqref{eq:d_ntn}. The CubeSat connects to relay $n$ during $[0, T_{\text{NTN}}/2]$ and switches to relay $n+1$ during $(T_{\text{NTN}}/2, T_{\text{NTN}}]$ as it becomes the closer node. Due to the circular symmetry of the constellation, the distance function in the second interval is a mirrored and shifted version of the first, a pattern that repeats periodically.
\begin{figure*}[!b]
\centering
\begin{minipage}{0.9\textwidth}
\begin{equation}
\label{eq:d_ntn}
d_{\text{NTN}}(t) =
\begin{cases}
\sqrt{R_{\text{CS}}^2 + R_{\text{NTN}}^2 - 2 R_{\text{CS}} R_{\text{NTN}} \cos(\omega_{\text{CS}} - \omega_{\text{NTN}})t }, & 0 < t < \frac{T_{\text{NTN}}}{2} \\[10pt]
\sqrt{R_{\text{CS}}^2 + R_{\text{NTN}}^2 - 2 R_{\text{CS}} R_{\text{NTN}} \cos(\omega_{\text{CS}} - \omega_{\text{NTN}})(T_{\text{NTN}} - t)}, & \frac{T_{\text{NTN}}}{2} < t < T_{\text{NTN}}
\end{cases}
\end{equation}
\end{minipage}
\end{figure*}
Equations \eqref{eq:dmin} and \eqref{eq:dmax} define the link distance bounds for \eqref{eq:d_ntn}. The minimum, $d^{\min}_{\text{NTN}}$, occurs when the CubeSat and relay are radially aligned, while the maximum, $d^{\max}_{\text{NTN}}$, occurs when the CubeSat is equidistant between two relays. These limits are essential for subsequent visibility and capacity evaluations.
\begin{equation}
\label{eq:dmin}
d^{\min}_{\text{NTN}} = |h_{\text{NTN}} - h_{\text{CS}}|
\end{equation}
\begin{equation}
\label{eq:dmax}
d^{\max}_{\text{NTN}} = \sqrt{R_{\text{CS}}^2 + R_{\text{NTN}}^2 - 2R_{\text{CS}}R_{\text{NTN}} \cos\left(\frac{\pi}{N_{\text{S}}}\right)}
\end{equation}

These limits analyze $d_{\mathrm{NTN}}(t)$ regardless of Earth's blockage, though \gls{los} is not always guaranteed. For example, if relay density is low, certain orbital positions may lack coverage. To account for this, we define the maximum distance for unobstructed \gls{los} as:
\begin{align}\label{eq:dB}
   d^{\text{B}}_{\text{NTN}} &= \bigg(h_{\text{CS}}^2 + h_{\text{NTN}}^2 + 2R_{\text{E}}(h_{\text{CS}} + h_{\text{NTN}}) +\\ \nonumber
   &+ 2\sqrt{h_{\text{CS}} h_{\text{NTN}} (2R_{\text{E}} + h_{\text{CS}}) (2R_{\text{E}} + h_{\text{NTN}})}\bigg)^{1/2}.
\end{align}

This distance represents the segment between the CubeSat and an \gls{ntn} relay when tangent to the Earth's surface. As shown in~\eqref{eq:dB}, the blockage distance depends primarily on the orbital altitudes $h_{\text{CS}}$ and $h_{\text{NTN}}$. Similar to the \gls{gs} case, we define a visibility function using $d_{\text{NTN}}^{\min}$, $d_{\text{NTN}}^{\max}$, and $d_{\text{NTN}}^{\text{B}}$:
\begin{equation}
\label{eq:vdntn}
V_{\text{NTN}}(d_{\text{NTN}}) =
\begin{cases}
1, & \hspace{-2mm}d_{\text{NTN}}^{\min} < d_{\text{NTN}} < \min\{d^{\text{B}}_{\text{NTN}}, d_{\text{NTN}}^{\max}\} \\
0, &\hspace{-2mm} \text{otherwise}.
\end{cases}
\end{equation}

Incorporating the time-dependence of $d_{\text{NTN}}(t)$ into \eqref{eq:vdntn} yields a visibility function that depends directly on time: 
\begin{equation}
\label{eq:vtntn}
V_{\text{NTN}}(t) =
\begin{cases}
0, & \frac{N_{\text{S}}\alpha}{2\pi} T_{\text{NTN}} < t < (1-\frac{N_{\text{S}}\alpha}{2\pi})T_{\text{NTN}}\\ 
1, & \text{Otherwise},
\end{cases}
\end{equation}

The parameter $\alpha$ captures the dependency on the number of relays and the altitudes of both the CubeSat and \gls{ntn} relays:
\begin{equation}
\label{eq:alpha}
\alpha=\cos^{-1} \left( \frac{R_{\text{CS}}^2 + R_{\text{NTN}}^2 - \left(\min\{d^{\text{B}}_{\text{NTN}}, d_{\text{NTN}}^{\max}\}\right)^2}{2 R_{\text{CS}} R_{\text{NTN}}} \right).
\end{equation}

In \eqref{eq:vtntn}, if $N_{\text{S}} \geq \lceil \frac{\pi}{\alpha} \rceil$, the first condition is impossible, rendering the visibility function equal to 1 for all $t$.

Ensuring permanent CubeSat coverage requires $d_{\text{NTN}}^{\max} < d^{\text{B}}_{\text{NTN}}$, meaning the maximum distance to any relay must remain below the Earth's blockage threshold. This yields the minimum number of \gls{ntn} relays required for continuous coverage:

\begin{equation}
\label{eq:n247}
N_{\text{S}}^* \hspace{-1mm}= \hspace{-1mm}\Biggl\lceil\hspace{-1mm}  
\frac{\pi}{\cos^{-1} \left(
\frac{R_{\text{E}}^2}{R_{\text{NTN}}R_{\text{CS}}} - \sqrt{\left(1 - \frac{R_{\text{E}}^2}{R_{\text{NTN}}^2}\right)\left(1 -\frac{R_{\text{E}}^2}{R_{\text{CS}}^2}\right)}
\right)} \hspace{-1mm} \Biggr\rceil 
\end{equation}

We study the statistical properties of $d_{\text{NTN}}$ using \eqref{eq:d_ntn} and assuming again a uniform distribution of time, i.e. $t\sim U[0, T_{\text{NTN}}]$. The resulting \gls{cdf} is reported in \eqref{eq:Fdntn}, with the corresponding derivations included in Appendix~\ref{app:CDF_derivation}.
\begin{figure*}[!b]
\centering
\vspace{-3mm}
\rule{\textwidth}{0.4pt} 
\begin{minipage}{0.9\textwidth}
\vspace{2mm}
\begin{equation}
\label{eq:Fdntn}
F_{d_{\text{NTN}}}(d) =
\begin{cases}
0, & d \leq d_{\text{NTN}}^{\min} \\
\frac{N_{\text{S}}}{\pi} \cos^{-1} \left(\frac{R_{\mathrm{CS}}^2 + R_{\mathrm{NTN}}^2 - d^2}{2 R_{\mathrm{CS}}R_{\mathrm{NTN}}}\right), & d_{\text{NTN}}^{\min} < d < \max\{d^{\text{B}}_{\text{NTN}}, d_{\text{NTN}}^{\max}\} \\
1, & d \geq \max\{d^{\text{B}}_{\text{NTN}}, d_{\text{NTN}}^{\max}\}.
\end{cases}
\end{equation}
\end{minipage}
\end{figure*}

The statistical models for $d_{\mathrm{GS}}$ and $d_{\mathrm{NTN}}$ provide a basis for evaluating distance-dependent performance metrics and their distributions. These models enable the analysis and comparison of different architectures, including ground-based and NTN-assisted scenarios. The derived distributions are used in subsequent sections to compute key performance metrics.

\subsection{Contact Probability}
As introduced in Section~\ref{subsec:metrics}, the contact probability $Q$ represents the ratio of time the CubeSat maintains \gls{los} with a receiving node and the scenario repetition period $T$. For the CubeSat-to-GS links, the scenario repetition period $T_{\text{GS}}$ corresponds to the least common multiple (LCM) of the Earth's rotation period, $T_{\mathrm{day}}$, and the CubeSat orbital period, $T_{\mathrm{CS}} = 1/\omega_{\mathrm{CS}}$, as follows:
\begin{equation}
\label{eq:Tdgs}
T_{\text{GS}} = T_{\text{res}} \mathrm{LCM}\left( \left\lfloor \frac{T_{\mathrm{day}}}{T_{\text{res}}} \right\rfloor, \left\lfloor \frac{T_{\mathrm{CS}}}{T_{\text{res}}} \right\rfloor \right),
\end{equation}
for which we use a temporal resolution $T_{\mathrm{res}}$ set to 1~min. 

By integrating the visibility function in~\eqref{eq:vdgs} over a full cycle and dividing by $T_{\text{GS}}$, we obtain the contact probability expressed in \eqref{eq:qgs}.
\begin{equation}
\label{eq:qgs}
Q_{\text{GS}} =\frac{\int_{0}^{T_{\text{GS}}} V_{\text{GS}}(t)\,dt
}{T_{\text{GS}}}= F_{d_{\text{GS}}}^* (d_{\text{GS}}^{\text{B}})
\end{equation}

Similarly, the contact probability $Q_{\mathrm{NTN}}$ is the ratio of contact time to the \gls{ntn} scenario periodicity, $T_{\text{NTN}}$, derived in~\eqref{eq:Tdntn}. By integrating the visibility function in~\eqref{eq:vtntn} over a full cycle, we determine the contact time as follows:
\begin{align}
\label{eq:qntn}
Q_{\text{NTN}} &=\frac{\int_{0}^{T_{\text{NTN}}} V_{\text{NTN}}(t)\,dt
}{T_{\text{NTN}}} \nonumber \\  &=\frac{\int_{0}^{\frac{N_{\text{S}}\alpha}{2\pi}T_{\text{NTN}}} 1\,dt\,+\int_{(1-\frac{N_{\text{S}}\alpha}{2\pi})T_{\text{NTN}}}^{T_{\text{NTN}}} 1\,dt
}{T_{\text{NTN}}}=\frac{N_{\text{S}}\alpha}{\pi}.
\end{align} 

As expected,, the contact probability $Q_{\text{NTN}}$ depends on the number of relays $N_{\text{S}}$ and orbital altitudes through $\alpha$ \eqref{eq:alpha}. If $N_{\text{S}} < N^*_{\text{S}}$, \gls{los} availability is intermittent due to geometric constraints, resulting in $Q_{\text{NTN}} < 1$. Conversely, if $N_{\text{S}} \geq N^*_{\text{S}}$, the CubeSat maintains continuous \gls{los}, and $Q_{\mathrm{NTN}} = 1$.

\subsection{Channel capacity}
\label{sec:channel_capacity}
The channel capacity $C(t)$ is inherently time-dependent due to the motion-driven variation of $d(t)$. While $P_{\mathrm{N}}$ remains constant—determined by bandwidth and receiver noise temperature—$P_{\mathrm{Rx}}(t)$ varies with the orbital geometry, making $C(t)$ highly sensitive to the satellites' positions. By substituting \eqref{eq:rxPower} into \eqref{eq:capacity_general}, the channel capacity for both GS and NTN cases is derived as:
\begin{equation}
\label{eq:channel_capacity}
\begin{aligned}
C(t) = W \log_2 \left( 
1 + \frac{P_{\mathrm{Tx}}G_{\text{CS}}G_{\text{Rx}}}{P_{\mathrm{N}}L(f_c, d(t))}
\right)V(t) \\ 
= W \log_2 \left( 
1 + \frac{b}{L_{\mathrm{abs}}(f_c, d(t))d^2(t)}
\right)V(t),
\end{aligned}
\end{equation}
where $b$ encapsulates the effects of transmit power, noise power, carrier frequency, and transmitter and receiver antenna gain as follows: 
\begin{equation}
\label{eq:u}
b = \frac{P_{\mathrm{Tx}}}{P_{\mathrm{N}}}G_{\text{CS}}G_{\text{Rx}}\left(\frac{\varsigma}{4 \pi f_c}\right)^2.
\end{equation}

Unlike the GS case, molecular absorption is negligible outside the atmosphere ($L_{\mathrm{abs}}(f_c,d)\approx1$). Consequently, for the NTN case, \eqref{eq:channel_capacity} simplifies to:
\begin{equation}
\label{eq:channel_capacity_ntn}
\begin{aligned}
C_{\text{NTN}}(t)  
= W \log_2 \left( 
1 + \frac{b}{d^2_{\text{NTN}}(t)}
\right)V_{\mathrm{NTN}}(t).
\end{aligned}
\end{equation}

Using the statistical characterization of $d(t)$, we derive the univariate distribution of $C(t)$. Specifically, the \gls{cdf} of channel capacity is determined using the visibility functions from Sec.~\ref{subsec:dist2cubesat} as follows:
\begin{equation}
\begin{split}
    F_C(c) &= P(C \leq c) \\
    &= P(C \leq c \mid V(d)=0)P(V(d)=0) \\
    &\quad + P(C \leq c \mid V(d)=1)P(V(d)=1),
\end{split}
\end{equation}
where $ c\geq 0$. Given that $C = 0$ during \gls{nlos} events, $P(C \leq c \mid V(d)=0)=1$. The probabilities of \gls{nlos} and \gls{los} are $1-Q$ and $Q$, respectively, where $Q$ is the contact probability. Applying these to~\eqref{eq:channel_capacity}, we obtain:
\begin{equation}\label{eq:F_C}
\begin{split}
    &\hspace{-3mm}F_C(c)= \\
    &\hspace{-3mm}1-Q P\hspace{-1mm}\left(\hspace{-1mm}d\leq\hspace{-1mm}\left(\frac{b}{L_{\text{abs}}(f_c, d)(2^{c/W}-1)}\right)^{0.5}\bigg|V(d)=1\hspace{-1mm}\right).
    \end{split}
\end{equation}

Substituting~\eqref{eq:Fdgs} and~\eqref{eq:Fdntn} into~\eqref{eq:F_C} yields the capacity \glspl{cdf} for \gls{gs} and \gls{ntn} in~\eqref{eq:F_C_GS} and \eqref{eq:F_C_NTN}. For brevity, the frequency and distance dependence of absorption loss is omitted in~\eqref{eq:F_C_GS}. These formulations enable quantifying the impact of relay density and altitude on throughput, as explored in Sec.~\ref{sec:results}.
 
\subsection{Total download capacity}\label{subsec:totalDcapacity}
Average system performance is evaluated through the total download capacity $\Gamma$, introduced in~\eqref{downloadcap}. This metric is obtained by integrating $C(t)$ over the scenario’s repetition period and scaling to a 24-hour duration. Substituting~\eqref{eq:channel_capacity} into~\eqref{downloadcap}, $\Gamma$ is calculated as:
\begin{equation}
\label{eq:download_capacity}
\hspace{-2mm}\Gamma =\frac{T_{\text{day}}}{T} \hspace{-1mm} \int_0^{T} \hspace{-1mm}W \log_2 \left( 1 + \frac{b}{L_{\mathrm{abs}}(f_c, d(t))d^2(t)} \right)V(t) dt
\end{equation}

For the NTN case, \eqref{eq:download_capacity} can be further developed by neglecting absorption loss and leveraging the symmetry of $d_{\text{NTN}}(t)$ around $T_{\text{NTN}}/2$. By substituting~\eqref{eq:Tdntn}, \eqref{eq:d_ntn}, and~\eqref{eq:vtntn} and applying a change of variables, we obtain:
\begin{equation}
\label{eq:download_capacity_ntn}
\begin{split}
    &\Gamma_{\text{NTN}} = \frac{T_{\text{day}}W}{\pi} \\
    & \int_0^{N_{\text{S}}\alpha} \hspace{-5mm}  \log_2 \left( 1 + \frac{b}{R_{\text{CS}}^{2} + R_{\text{NTN}}^{2} - 2R_{\text{CS}}R_{\text{NTN}}\cos(\frac{\phi}{N_{\text{S}}})}
    \right)d\phi.
    \end{split}
\end{equation}
Since the integrals for the \gls{ntn} and \gls{gs} cases lack analytical solutions, they must be computed numerically.

While \eqref{eq:download_capacity_ntn} shows that total download capacity increases monotonically with the number of relays, we must determine if this growth is bounded. To evaluate whether capacity increases indefinitely with $N_{\text{S}}$, we examine the limit:
\vspace{-2mm}
\begin{equation}\label{eq:gamma_n2infinity}
    \lim_{N \to \infty} \Gamma_{\mathrm{NTN}}=\frac{W}{T_{\text{day}}}\log_2\left(1+\frac{b}{(d_{\mathrm{NTN}}^{\min})^2}\right).
\end{equation}
\vspace{-3mm}

This result reveals a fundamental bound: the total download capacity cannot be arbitrarily increased by adding more satellites. Instead, \emph{the download capacity is bounded by the channel capacity at the minimum distance between the CubeSat and the \gls{ntn} relays}, $d_{\mathrm{NTN}}^{\min}$. This is an important theoretical insight from our work that is further elaborated on in Sec.~\ref{sec:results}.

\subsection{Energy efficiency}
\label{sec:energy_efficiency}
Using the general definition in~\eqref{eq:energyeff_general}, we compute the energy efficiency for both cases by substituting the contact time $T_c = QT$ and the total download capacity over one period. Thus, ~\eqref{eq:energyeff_general} simplifies to:
\vspace{-1mm}
\begin{equation}\label{eq:energyeff}
    \eta = \frac{\Gamma}{P_{\text{Tx}}QT_{\text{day}}}.
\end{equation}
\vspace{-2mm}

Accordingly, the energy efficiency expressions for the GS and NTN cases are obtained from~\eqref{eq:energyeff} by using the respective values of $\Gamma$, $T$, and $Q$ previously derived.
 \begin{figure*}[!b]
\centering
\vspace{-5mm}
\rule{\textwidth}{0.4pt} 
\begin{minipage}{0.9\textwidth}
\vspace{-2mm}
\begin{equation}\label{eq:F_C_GS}
    F_{C_{\text{GS}}}(c)=\begin{cases}
        0, & c<0\\
        1-Q_{\text{GS}}, & 0\leq c \leq W\log_2\left(1+\frac{b}{L_{\mathrm{abs}}(d_{\text{GS}}^{B})^2}\right)\\
        1-Q_{\text{GS}}F^*_{d_{\text{GS}}}\left(\sqrt{\frac{b}{L_{\text{abs}}(2^{c/W}-1)}}\right), & W\log_2\left(1+\frac{b}{L_{\mathrm{abs}}(d_{\text{GS}}^{B})^2}\right)<c<W\log_2\left(1+\frac{b}{L_{\mathrm{abs}}(d_{\text{GS}}^{\min})^2}\right)\\
1, & c\geq W\log_2\left(1+\frac{b}{L_{\mathrm{abs}}(d_{\text{GS}}^{\min})^2}\right)
    \end{cases}
\end{equation}
\end{minipage}
\end{figure*}
\begin{figure*}[!b]
\centering
\begin{minipage}{0.9\textwidth}
\begin{equation}\label{eq:F_C_NTN}
    \hspace{-5mm}F_{C_{\text{NTN}}}(c)\hspace{-1mm}=\hspace{-1mm}\begin{cases}
        0, &\hspace{-3mm} c<0\\
        
        1-Q_{\text{NTN}}, &\hspace{-3mm} 0\leq c \leq W\log_2\left(1+\frac{b}{(\min\{d^{\text{B}}_{\text{NTN}}, d_{\text{NTN}}^{\max}\})^2}\right)\\
        
        1-Q_{\text{NTN}}\frac{N_{\text{S}}}{\pi}\cos^{-1}\hspace{-1mm}\left(\frac{R_{\mathrm{CS}}^2 + R_{\mathrm{NTN}}^2 - \frac{b}{2^{\frac{c}{W}} - 1}}
{2 R_{\mathrm{CS}}R_{\mathrm{NTN}}}\right)\hspace{-1mm}, &\hspace{-3mm} W\log_2\hspace{-1mm}\left(1\hspace{-1mm}+\hspace{-1mm}\frac{b}{(\min\{d^{\text{B}}_{\text{NTN}},d_{\text{NTN}}^{\max}\})^2}\right)<c<W\log_2\hspace{-1mm}\left(1\hspace{-1mm}+\hspace{-1mm}\frac{b}{(d_{\text{NTN}}^{\min})^2}\hspace{-1mm}\right)\\

1, &\hspace{-3mm} c\geq W\log_2\left(1+\frac{b}{(d_{\text{NTN}}^{\min})^2}\right)
    \end{cases}
\end{equation}
\end{minipage}
\end{figure*}

\section{Numerical Results}\label{sec:results}
In this section, the mathematical results from Section~\ref{sec:analysis} are numerically elaborated. We particularly compare the system performance dynamics for four modeled CubeSat connectivity scenarios (see Section~\ref{sec:system_model}), contrasting the existing \gls{gs}-centric and alternative \gls{ntn}-assisted configurations.
\subsubsection{Ground-Station Configuration}
The single-GS scenario uses a station in Svalbard, whose near-polar latitude provides excellent \gls{sso} coverage. For the multi-GS configuration, we select 17 locations from the Leaf Space "Leaf Line" network—a global \gls{gsaas} provider widely used in small-satellite missions~\cite{LeafLine}. This architecture provides operationally validated worldwide coverage aligned with current mission practices.

\subsubsection{CubeSat and NTN Satellites Orbital Configuration}
The CubeSat follows a \gls{sso}, typical for missions requiring consistent illumination and predictable revisits. The co-planar \gls{ntn} relays are positioned in circular \gls{leo} orbits at $h_{\mathrm{NTN}} = 550$~km. This co-planar assumption is physically sound for \gls{sso} orbits, reflecting the high inclination of polar orbits in modern constellations~\cite{pachler2021updated}.

We evaluate three wireless connectivity options: Ku-band, Sub-THz, and Sub-THz Next-Generation (NG). The corresponding link-budget parameters for these three radio setups, based on state-of-the-art mmWave and sub-THz hardware, are summarized in Table~\ref{tab:LBparameters}. The difference between the two sub-THz options is that the former assumes state-of-the-art sub-THz hardware, while the latter assumes prospective better and higher-output components to become available soon~\cite{jornet_proc_ieee}. We assume antenna diameters of $10$~cm (CubeSat) and $60$~cm (NTN relay or GS), with an antenna temperature $T_{\text{a}}=300$~K~\cite{Maral2009satellite}. The aperture efficiency is 60\% for Ku and standard Sub-THz, and 95\% for Sub-THz NG, assuming such efficiency will soon be available through advanced precision manufacturing.
\begin{table}[h]
\vspace{-3mm}
    \centering
    \caption{Key radio technology parameters}
    \label{tab:LBparameters}
    \begin{tabular}{|l|c|c|c|}
        \hline
        \textbf{Parameter} & \textbf{Ku Band} & \textbf{sub-THz} & \textbf{sub-THz NG}\\
        \hline
        $f_c$    & 18 GHz \cite{Maral2009satellite}  & \multicolumn{2}{c|}{220 GHz \cite{Siles2018new}} \\ \hline
        $W$      & 400 MHz \cite{chen2024catalyzing} &  \multicolumn{2}{c|}{5 GHz \cite{Sen2021versatile}} \\ \hline
        $P_{\mathrm{Tx}}$ & 40 dBm \cite{chen2024catalyzing} & 20 dBm \cite{Siles2018new} &  27 dBm \cite{cooper2025power} \\ \hline
        $G_{\text{Tx}}$ & 23~dBi  & 45~dBi & 47~dBi \\ \hline
        $G_{\text{Rx}}$ & 39~dBi & 61~dBi & 63~dBi \\ \hline
        $F_{\text{N}}$      & 3 dB \cite{chen2024catalyzing}   &  \multicolumn{2}{c|}{7 dB \cite{Siles2018new}} \\ \hline
    \end{tabular}
\end{table}
\vspace{-2mm}

\subsection{Simulation setup}\label{subsec:simulationsetup}
Our mathematical results are cross-verified using the MATLAB Satellite Communications Toolbox. We primarily ensure that the analytically derived distances between all communicating nodes match simulation results, specifically addressing the complex scenario geometry and non-linear orbital dynamics described in Section~\ref{sec:analysis}. Following position and distance verification, propagation and link-budget computations are performed according to the methodology in Sec.~\ref{sec:system_model}.

\begin{figure}[!h]
    \centering
    \subfloat[$h_{\mathrm{CS}} = 400$ km]{
        \includegraphics[width=0.98\linewidth]{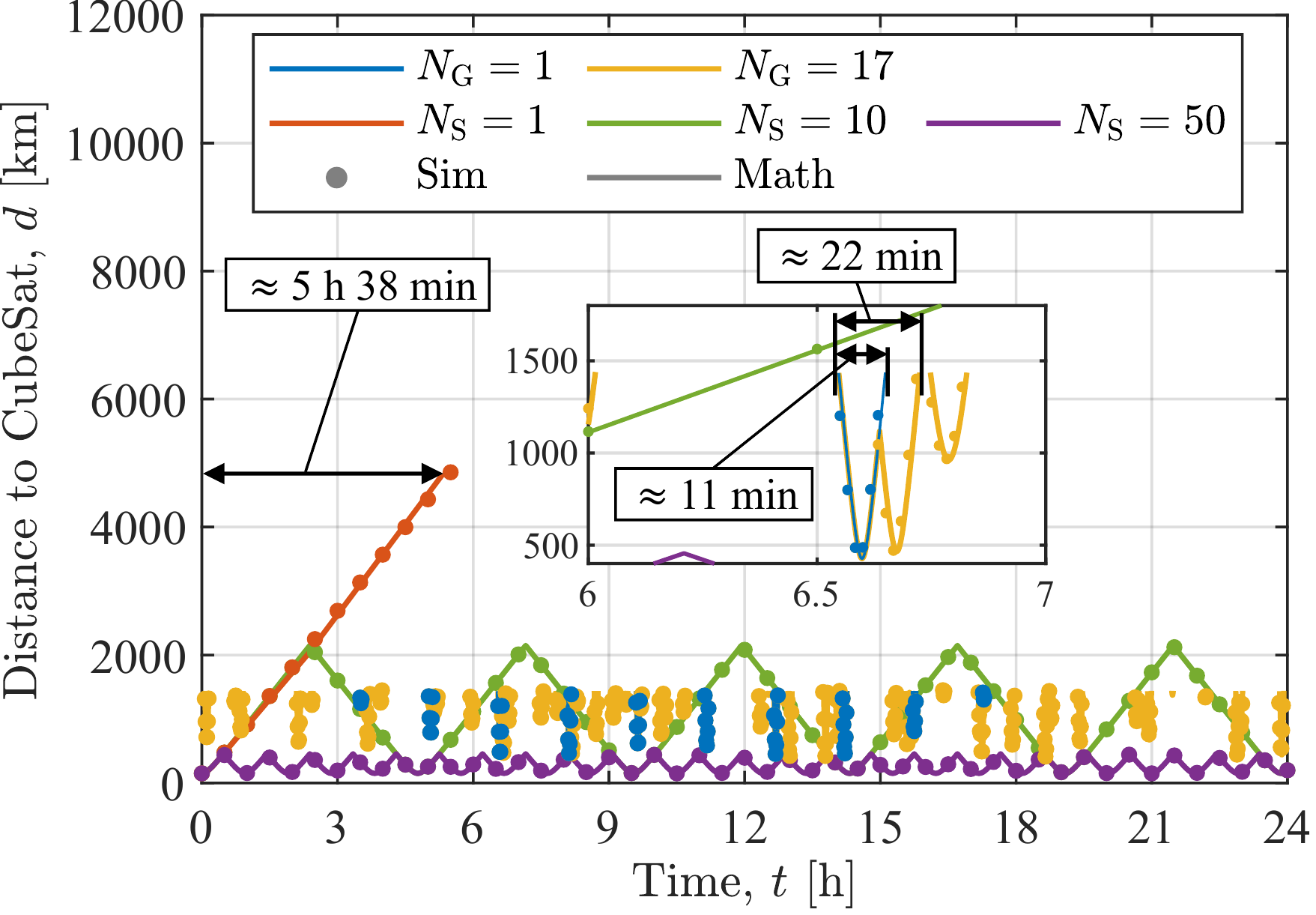}
        \label{fig:linkdistance_400}
    }\hfill
    \subfloat[$h_{\mathrm{CS}} = 2000$ km]{
        \includegraphics[width=0.98\linewidth]{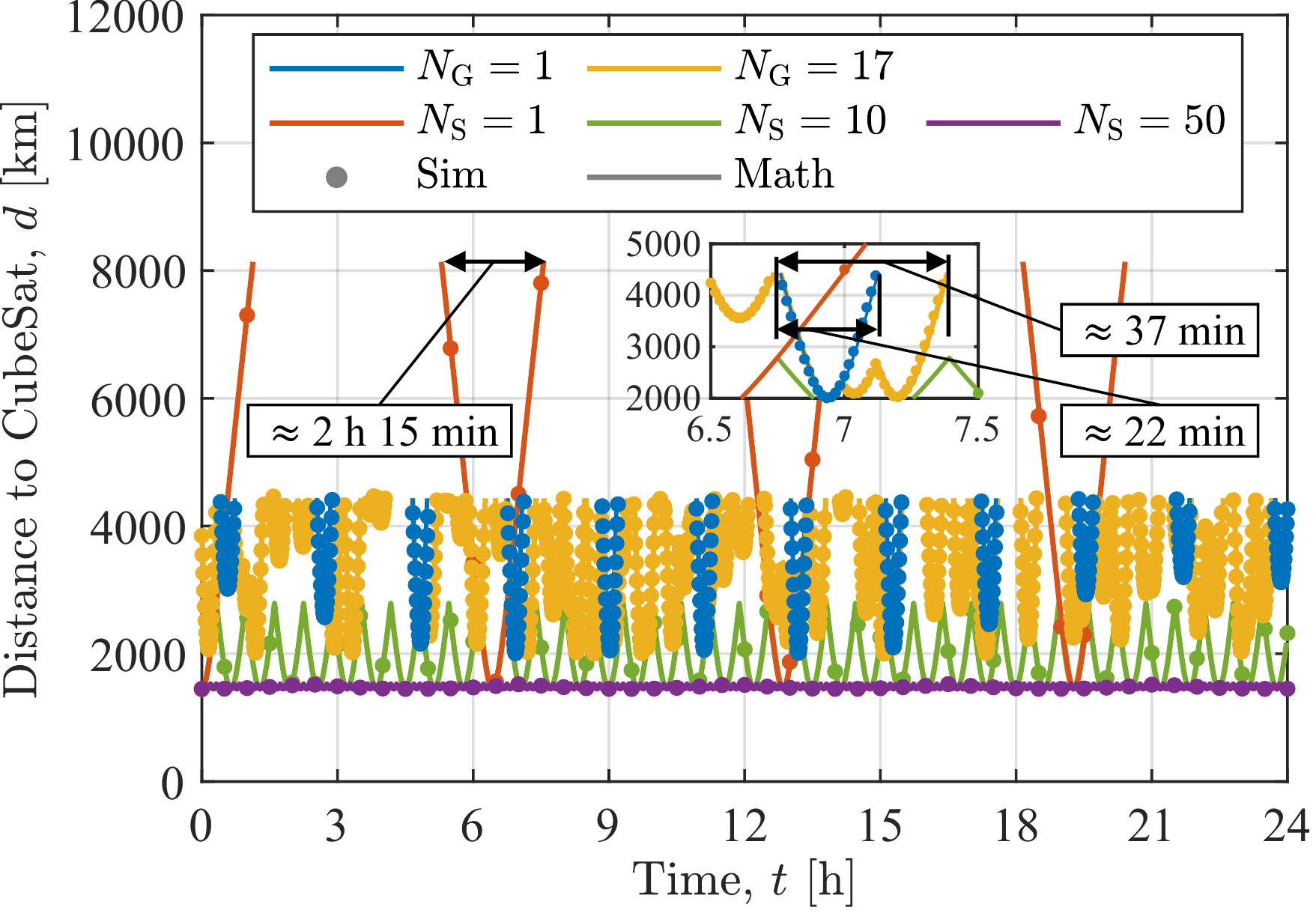}
        \label{fig:linkdistance_2000}
    }
    \caption{Link distance for different CubeSat altitudes $h_{\mathrm{CS}}$.}
    \vspace{-3mm}
    \label{fig:d_vs_time}
\end{figure}

\begin{figure}[!h]
\label{fig:linkdistance_CDF}
    \centering
    \subfloat[CubeSat-to-GS]{
        \includegraphics[width=0.98\linewidth]{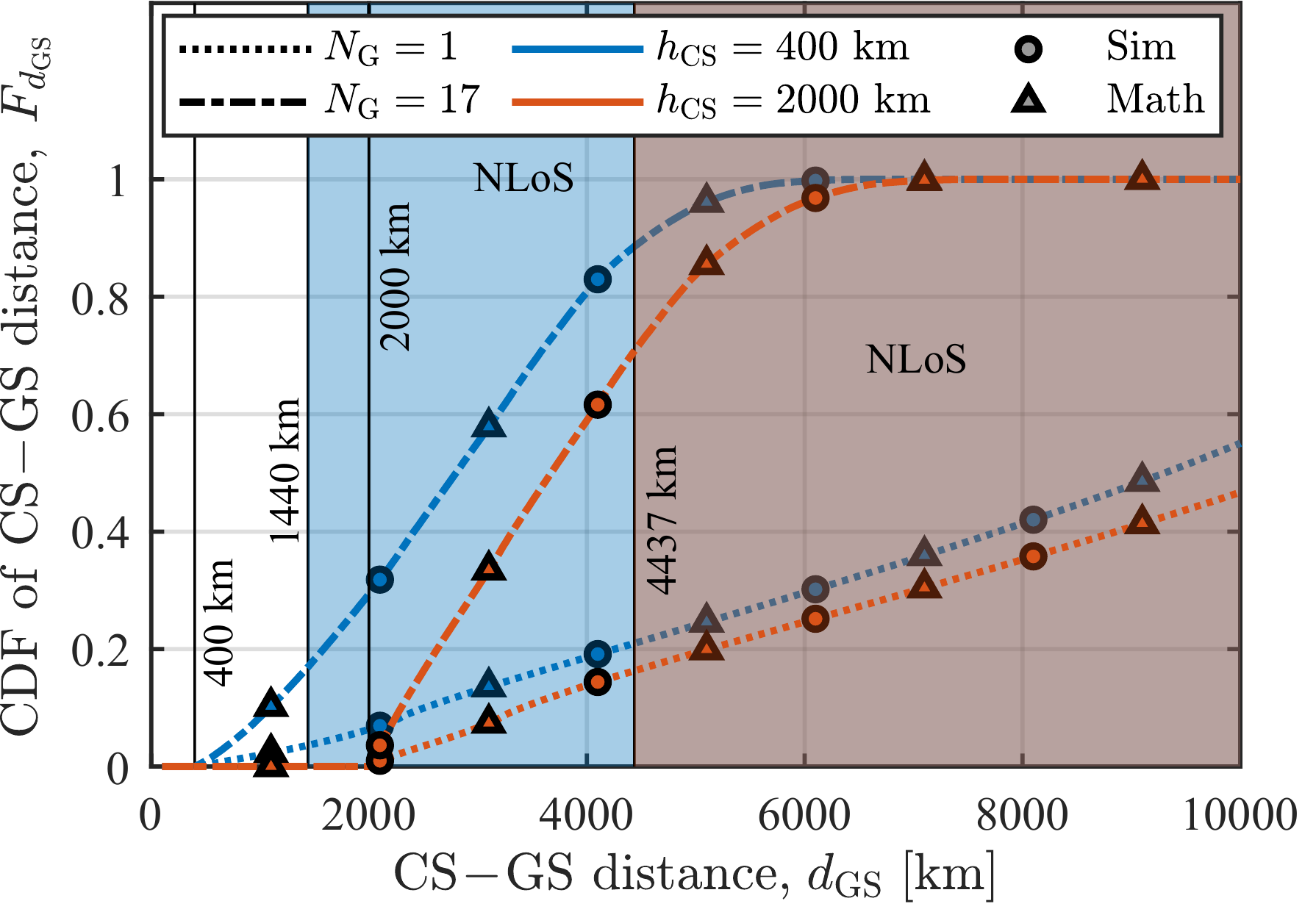}
        \label{fig:linkdistance_CDF_GS}
    }\hfill
    \subfloat[CubeSat-to-NTN]{
        \includegraphics[width=0.98\linewidth]{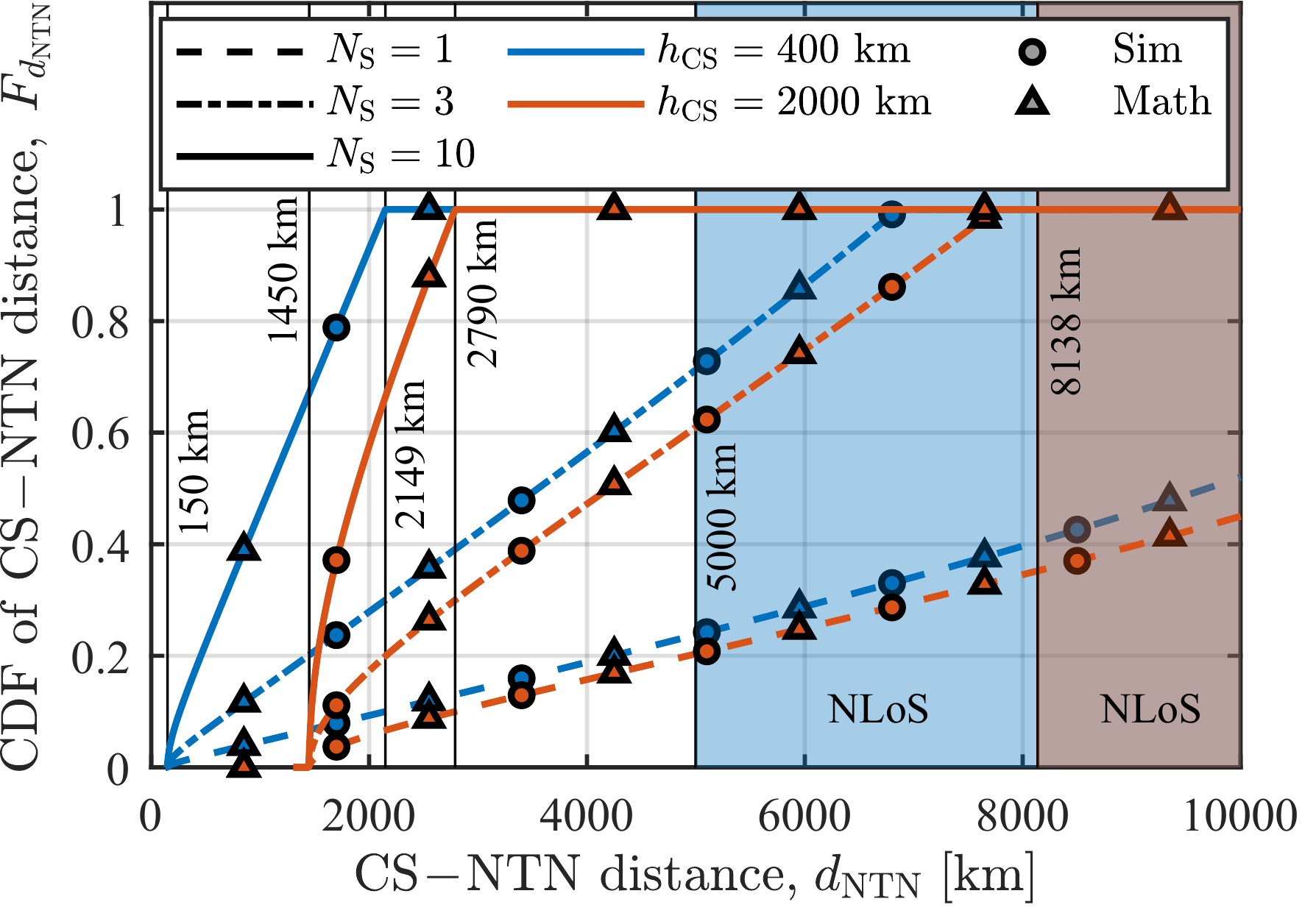}
        \label{fig:linkdistance_CDF_NTN}
    }
    \caption{Link distance CDF at distinct CubeSat altitudes $h_{\mathrm{CS}}$.}
    \vspace{-2mm}
    \label{fig:linkdistance_CDF}
\end{figure}

\subsection{Link distance analysis}\label{subsec:linkdistance_num}
This subsection evaluates the temporal evolution and statistical distribution of link distances for both \gls{ntn}-based and \gls{gs}-based scenarios. We compare results from the analytical framework against the simulation procedure to validate the link distance modeling for both scenarios.

\begin{figure}[!t]
    \vspace{-2mm}
    \centering
    \subfloat[Probability of contact.]{
        \includegraphics[height=5cm]{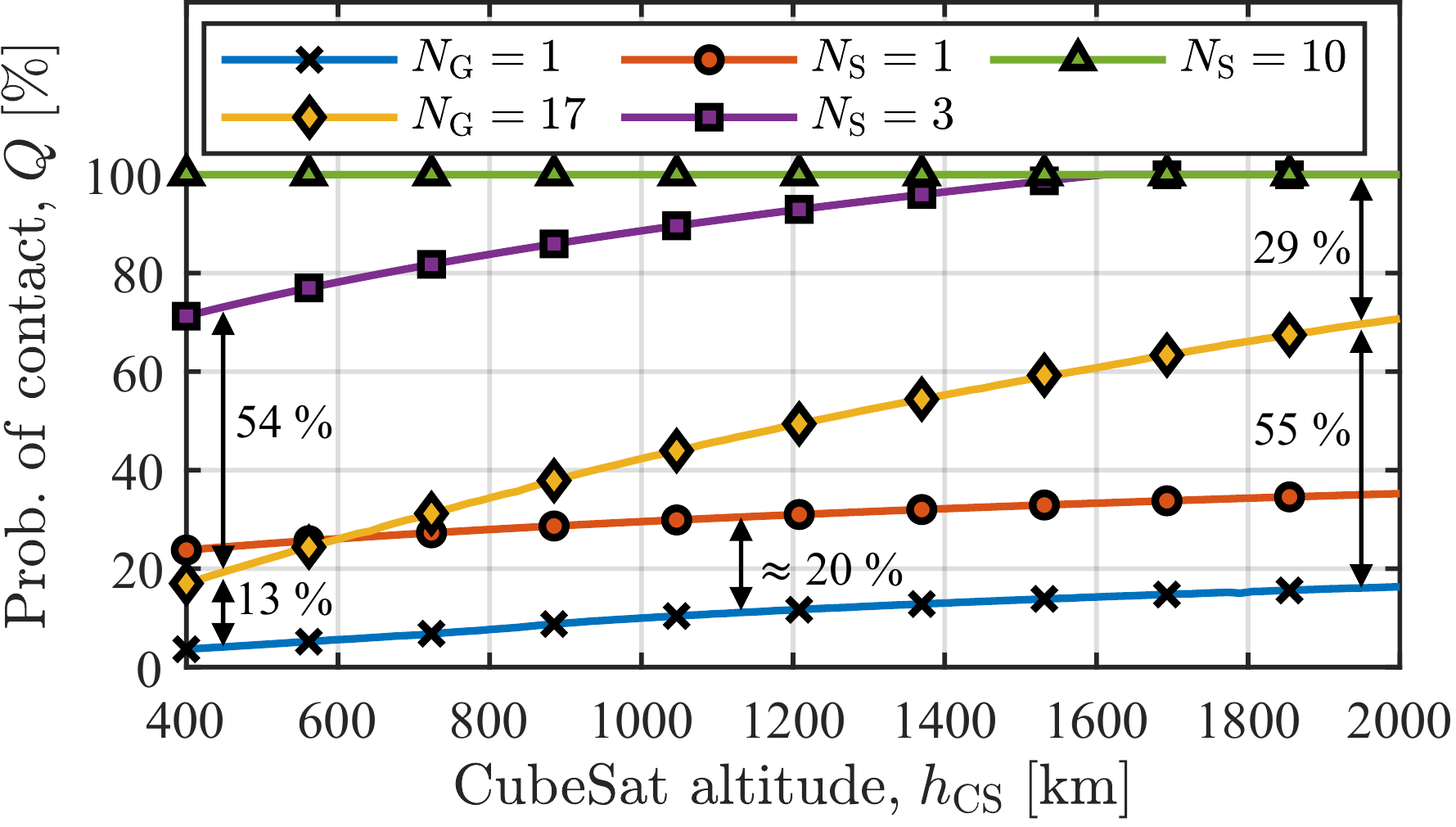}
        \label{fig:prob_of_contact}
    }\hfill
    \subfloat[Minimum number of relays for 24/7 contact.]{
        \includegraphics[height=5cm]{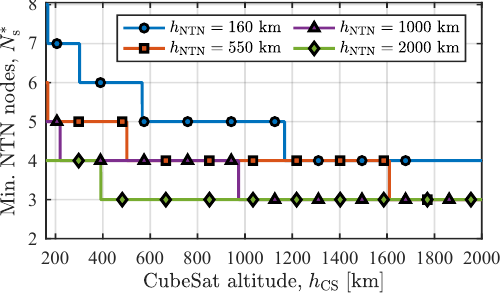}
        \label{fig:N247}
    }
    \caption{Probability of contact and minimum number of relays for 24/7 contact.}
    \vspace{-7mm}
    \label{fig:linkdistance_comparison}
\end{figure}

\begin{figure}[!b]
    \centering
    \includegraphics[width=0.98\linewidth]{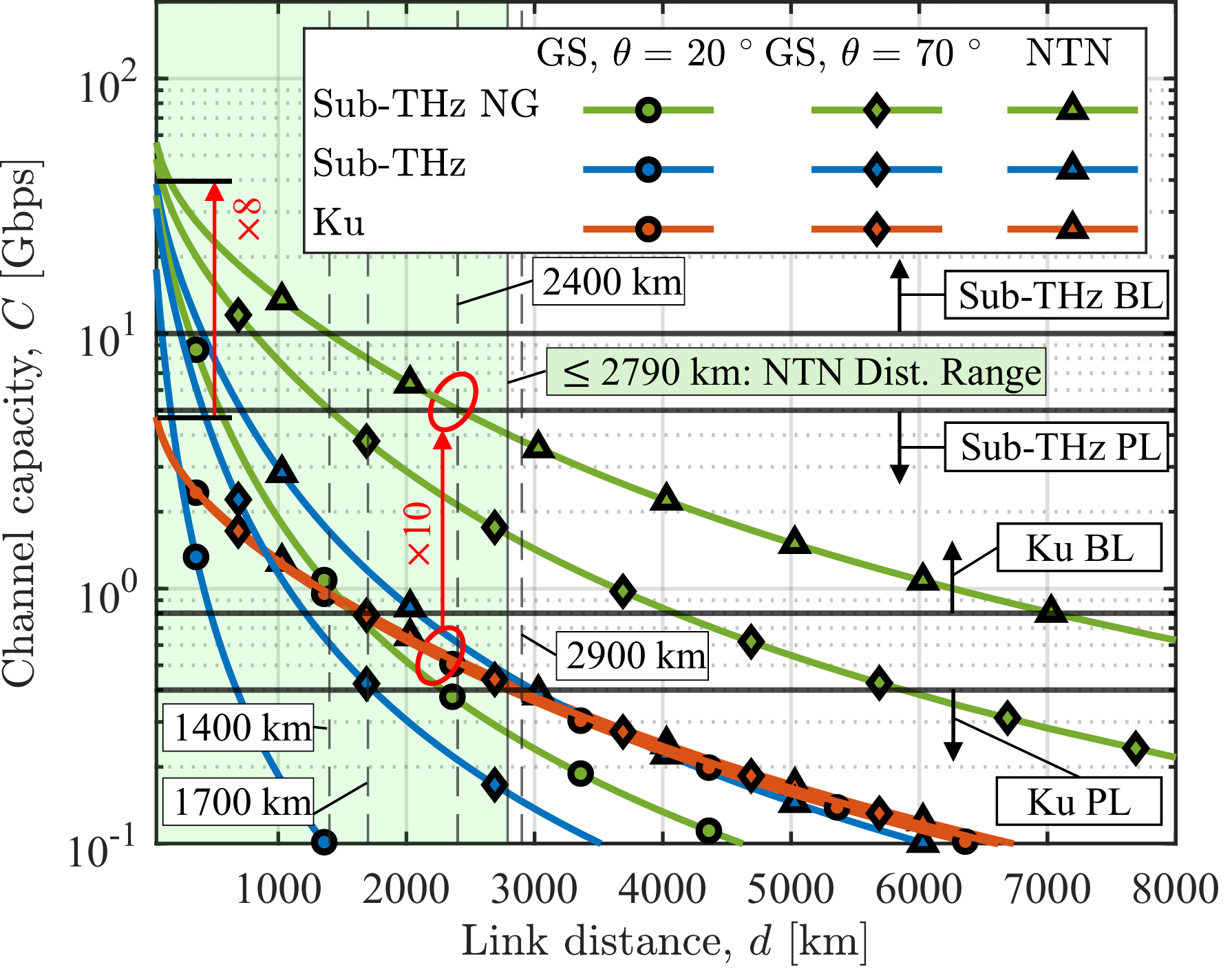}
    \caption{Chanel capacity offered by the different wireless technologies considered.}
    \label{fig:C_vs_d}
    \vspace{-5mm}
\end{figure}

Fig.~\ref{fig:d_vs_time} illustrates the link distance over time for CubeSat altitudes of $400$~km and $2000$~km. GS contacts are brief, typically between $11$ and $22$ minutes for a single station. While increasing the number of \glspl{gs} adds more contact events, continuous connectivity remains largely unavailable. In contrast, a single NTN relay provides significantly longer contact durations (approx. $2.25$ to $5.7$ hours), albeit often at greater distances than GS links. Continuous coverage is achieved with $N_{\text{S}}=10$ relays, and further increasing relay density reduces the average link distance, which converges toward the altitude difference ($|h_{\text{CS}} - h_{\text{NTN}}|$) at $N_{\text{S}}=50$.

Fig.~\ref{fig:linkdistance_CDF} validates the statistical link distance analysis by comparing analytical \glspl{cdf} from \eqref{eq:Fdgs} and \eqref{eq:Fdntn} with simulation results. Shaded areas denote \gls{nlos} regions, where the intersection of each curve with the boundary represents the contact probability $Q$. For GS links (Fig.~\ref{fig:linkdistance_CDF_GS}), $Q$ remains below 20\% for $N_{\mathrm{G}}=1$, and while a multi-GS network improves $Q$, continuous coverage is not reached.

In contrast, Fig.~\ref{fig:linkdistance_CDF_NTN} shows that higher CubeSat altitudes increase $Q$ at the expense of longer link distances. Notably, multiple relays enable continuous connectivity ($Q=1$), as the \glspl{cdf} saturate before the \gls{nlos} threshold. For $N_{\text{S}}=10$, the maximum distance remains below $2149$~km and $2790$~km for $400$~km and $2000$~km altitudes, respectively—well within the corresponding \gls{nlos} boundaries of $5000$~km and $8138$~km.

As shown in Figs.~\ref{fig:d_vs_time} and \ref{fig:linkdistance_CDF}, the analytical and simulated results match closely across all altitudes and \gls{ntn} configurations. This high correlation validates the geometric foundations of our mathematical model. Consequently, subsequent evaluations for channel capacity, total download capacity, and energy efficiency rely exclusively on the analytical framework.

\subsection{Probability of contact}\label{subsec:Q_num}

To better understand the dynamics of the probability of contact beyond Fig.~\ref{fig:linkdistance_CDF}, we plot this performance metric as a function of CubeSat orbital altitude in Fig.~\ref {fig:prob_of_contact}. As shown, the probability of contact $Q$ increases monotonically with CubeSat altitude, as higher orbits extend contact duration. While this increase is mild for single-node configurations, the single-relay \gls{ntn} consistently provides a $\approx 20\%$ higher $Q$ than a single \gls{gs}.

Expanding the \gls{gs} network to 17 stations improves $Q$ by $13\%$–$55\%$, yet performance remains inferior to even a small relay constellation. Notably, $N_{\text{S}}=3$ relays outperform the 17-node \gls{gs} architecture, while $N_{\text{S}}=10$ achieves $100\%$ contact probability across all altitudes. This demonstrates that continuous coverage is feasible with a modest \gls{ntn} deployment, a result practically unattainable with ground-based infrastructures.

Fig.~\ref{fig:N247} illustrates the minimum number of relays $N^*_{\text{S}}$ required for continuous connectivity, as derived in \eqref{eq:n247}. The requirement for $N^*_{\text{S}}$ decreases as either the \gls{ntn} altitude ($h_{\mathrm{NTN}}$) or CubeSat altitude ($h_{\mathrm{CS}}$) increases. For $h_{\mathrm{NTN}} = 550$~km, the relay requirement drops from five to three as the CubeSat altitude rises from $400$~km to $2000$~km. Notably, even at the very low \gls{leo} altitudes ($160$~km), continuous connectivity is achievable with fewer than ten relays.

\subsection{Channel capacity}\label{subsec:C_num}
This section evaluates the joint impact of orbital dynamics and wireless technologies on the available CubeSat channel capacity. Fig.~\ref{fig:C_vs_d} evaluates the impact of orbital dynamics and wireless technology on channel capacity, marking the boundaries for power-limited (PL, $C/B < 1$ bps/Hz) and bandwidth-limited (BL, $C/B > 2$ bps/Hz) regimes. While Ku-band performance is stable due to negligible absorption, Sub-THz capacity varies significantly, improving drastically when transitioning from absorption-prone, low-elevation \gls{gs} links to absorption-free \gls{ntn} scenarios. Current Sub-THz hardware (100 mW) matches or exceeds Ku-band capacity in \gls{ntn} configurations for distances below $3200$~km and achieves an eightfold gain at $50$~km.Sub-THz NG provides the most significant gains, maintaining a capacity approximately one order of magnitude higher than Ku-band across all distances. Both Sub-THz NG and Ku-band exhibit similar spectral efficiency transitions—entering the PL region above $\approx 2400$--$2900$~km and the BL region below $\approx 1400$--$1700$~km. This indicates that the Sub-THz capacity advantage is driven primarily by its significantly larger available bandwidth rather than superior spectral efficiency.

\begin{figure*}[!t]
    \centering
    \subfloat[$h_{\mathrm{CS}} = 400$ km]{
        \includegraphics[width=0.999\linewidth]{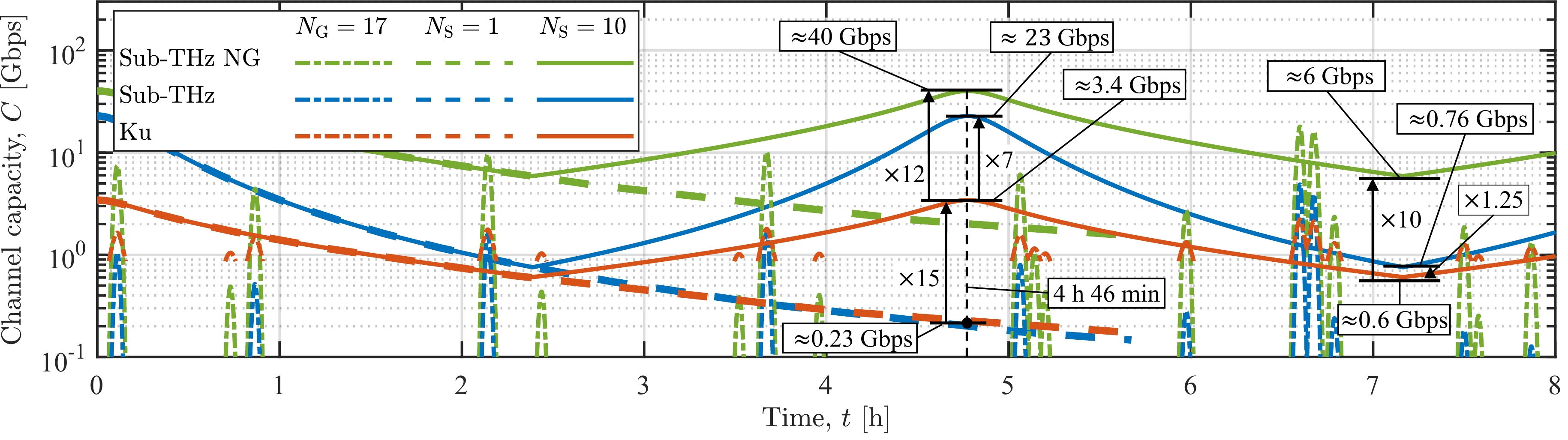}
        \label{fig:cap_h400}
    }\\[1ex]
    \subfloat[$h_{\mathrm{CS}} = 2000$ km]{
        \includegraphics[width=0.999\linewidth]{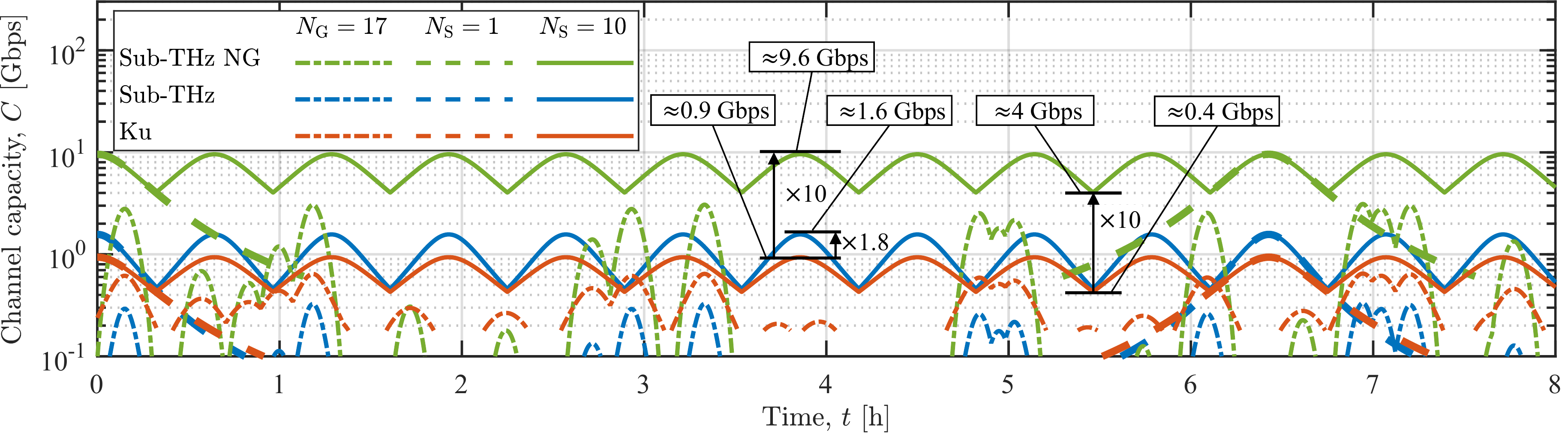}
        \label{fig:cap_h2000}
    }
    \caption{Channel capacity over time for different CubeSat orbital altitudes $h_{\mathrm{CS}}$.}
    \vspace{-5mm}
    \label{fig:cap_vs_t}
\end{figure*}

To capture the temporal dynamics of each scenario, Fig.~\ref{fig:cap_vs_t} illustrates the instantaneous channel capacity for $h_{\text{CS}}$ of $400$~km and $2000$~km. Higher altitudes provide longer contacts and more stable average capacities, though lower altitudes yield higher absolute peaks due to reduced path loss. Spatial diversity is critical; at $t = 4$~h $46$~min ($h_{\text{CS}} = 400$~km), $N_{\text{S}}=10$ relays maintain Ku capacity near $3.4$~Gbps, whereas a single node drops to $0.23$~Gbps. Sub-THz NG in \gls{gs} scenarios generally outperforms Ku-band but suffers from severe absorption at low elevation angles, causing rapid degradation away from the zenith. In contrast, Sub-THz \gls{ntn} architectures decouple high-frequency performance from atmospheric impairments, providing consistent gains—up to 7-fold for Sub-THz and 12-fold for Sub-THz NG in peak capacity. These results demonstrate that \gls{ntn} configurations are essential for achieving the reliable, multi-gigabit throughput required for data-intensive space applications.

\begin{figure}[!h]
    \centering
    \includegraphics[width=1\linewidth]{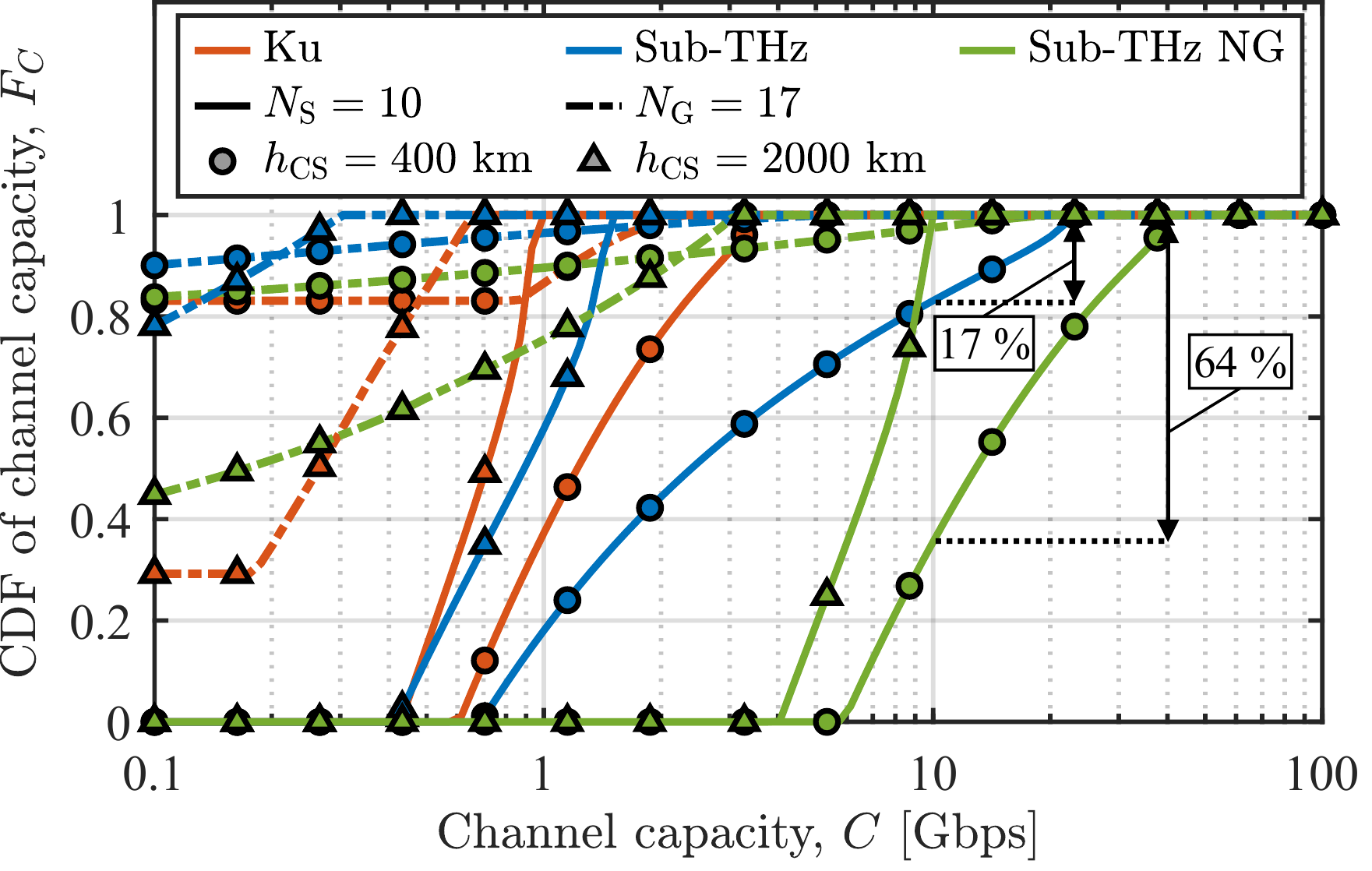}
    \caption{CDF of the channel capacity for the different architectures and technologies considered.}
    \label{fig:F_C}
    \vspace{-7mm}
\end{figure}

Fig.~\ref{fig:F_C} provides a statistical characterization of channel capacity for $N_{\mathrm{S}}=10$ and $N_{\mathrm{G}}=17$ configurations. For GS architectures, the offset $F_{\mathrm{C}}(0) = 1-Q$ represents outage periods where no contact exists. The results confirm that lower CubeSat altitudes ($400$~km) yield higher peak capacities and greater variability than higher orbits,  as lower altitude \gls{cdf} curves exhibit a broader capacity range and saturate at higher values compared to those for $2000$~km. Notably, all GS-based curves saturate well below NTN counterparts. Sub-THz NTNs emerge as the only viable solution for high-data-rate demands, with $P(C \geq 10 \text{ Gbps})$ reaching $17\%$ for standard Sub-THz and $64\%$ for Sub-THz NG. Conversely, while Ku-band GS systems remain suitable for applications below $1$~Gbps, they fail to achieve the $10$~Gbps threshold with meaningful probability.

\subsection{Total download capacity}\label{subsec:downloadcapacity_num}
\begin{figure}[!t]
    \vspace{-7mm}
    \centering
    \subfloat[Across different architectures and technologies]{
        \includegraphics[width=0.98\columnwidth]{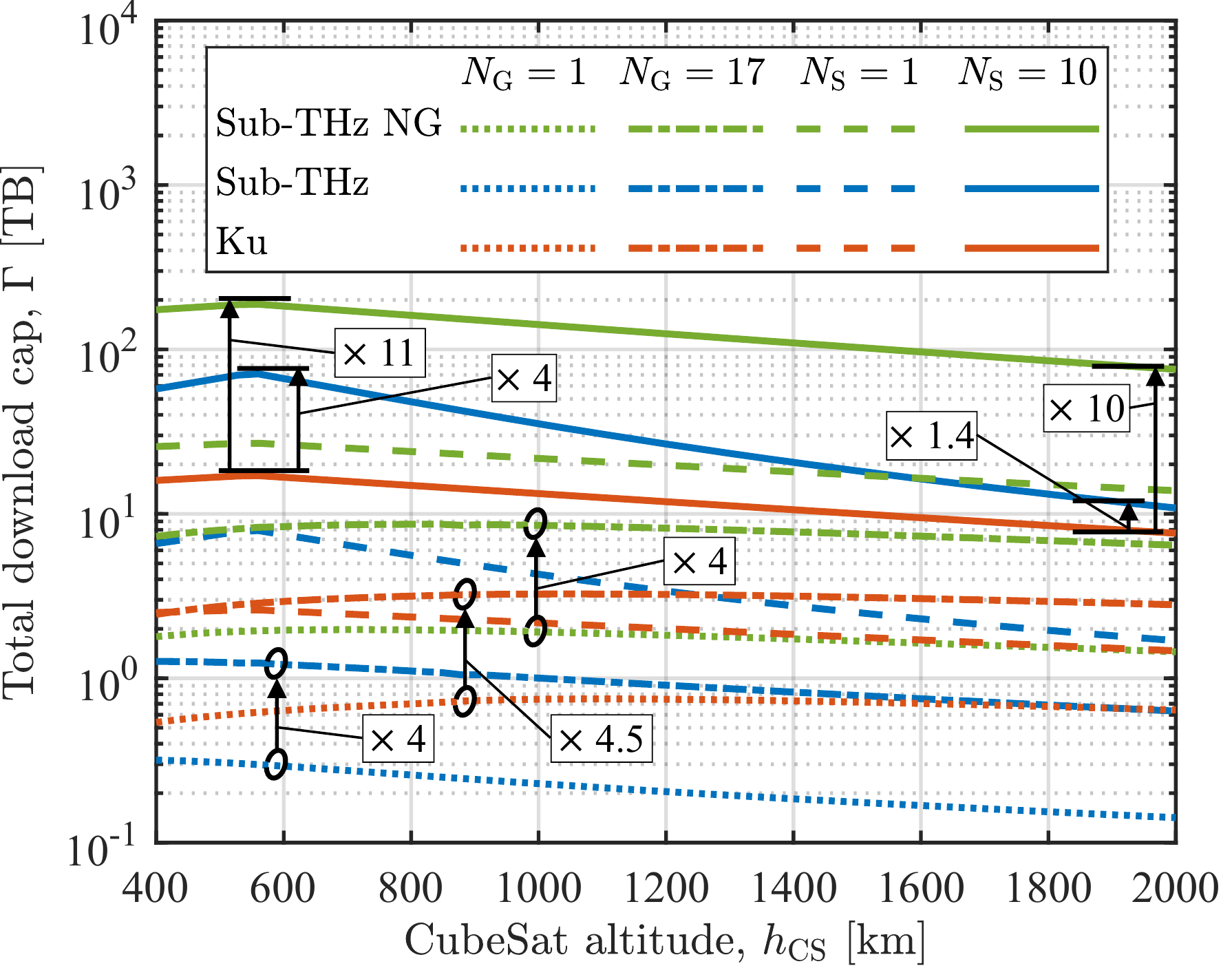}
        \label{fig:DownloadCapacity_vs_h}
    }\hfill
    \subfloat[As a function of the number of NTN satellites $N$]{
        \includegraphics[width=0.98\columnwidth]{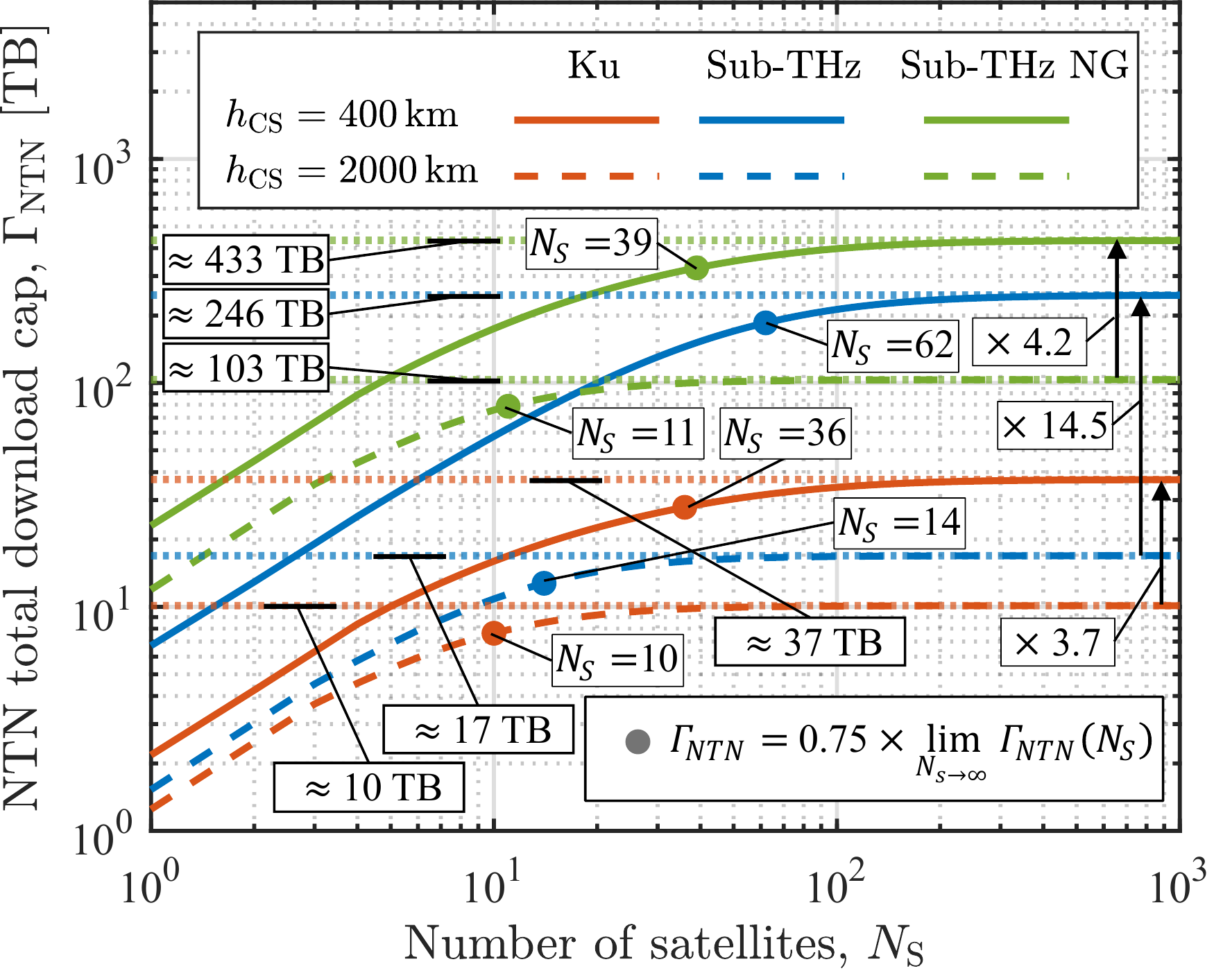}
        \label{fig:DownloadCapacity_vs_N}
    }
    \caption{Total download capacity $\Gamma$ over one day of coverage.}
    \vspace{-6mm}
    \label{fig:DownloadCapacity}
\end{figure}

To move beyond time-varying dynamics and evaluate the net throughput of the system, we analyze the 24-hour total download capacity $\Gamma$. As shown in Fig.~\ref{fig:DownloadCapacity}, baseline single-GS performance ranges from $140$--$750$~GB/day for Ku and Sub-THz, reaching $\approx 2$~TB/day for Sub-THz NG. Multi-GS configurations increase $\Gamma$ by a factor of $4$--$4.5$ across all technologies due to increased contact availability.

Notably, a single-GS Sub-THz NG setup or a single-node Ku \gls{ntn} achieves performance comparable to a 17-node Ku \gls{gs} network. This highlights the core potential of the proposed framework: both the transition to \gls{ntn}-based architectures and the adoption of Sub-THz technology yield substantial gains over state-of-the-art single-GS infrastructures.

The most significant performance increase occurs when combining \gls{ntn}-based architectures with Sub-THz radios, achieving more than an order of magnitude improvement in $\Gamma$ over Ku-band across all altitudes. Current Sub-THz technology yields gains of $\times1.4$ to $\times4$, while Sub-THz NG technology extends this to $\times10$--$\times11$. These results underscore that the maximum potential of \gls{ntn} architectures is realized by leveraging Sub-THz bands.

Notably, the curves in Fig.~\ref{fig:DownloadCapacity_vs_h} exhibit local maxima due to the trade-off between contact probability and link distance. While higher CubeSat altitudes increase the probability of contact (Fig.~\ref{fig:prob_of_contact}), they simultaneously increase link distances (Fig.~\ref{fig:linkdistance_CDF}). The interplay of these opposing effects yields a peak in $\Gamma$, which for \gls{ntn} architectures occurs at $550$~km—the altitude of the relay nodes.

As shown in Fig.~\ref{fig:DownloadCapacity_vs_N}, increasing the number of \gls{ntn} relays enhances $\Gamma$, yet \eqref{eq:gamma_n2infinity} and Fig.~\ref{fig:DownloadCapacity_vs_N} confirm this performance is bounded. These upper bounds are significantly higher for Sub-THz than for Ku-band. At $2000$~km, the maximum daily download capacities are $10$, $17$, and $103$~TB for Ku, Sub-THz, and Sub-THz NG, respectively. Reducing the altitude to $400$~km increases these bounds by $\approx\times4$ for Ku and Sub-THz NG, while standard Sub-THz experiences a $\times14.5$ gain as the link transitions from a PL to a BL regime. The rate at which $\Gamma$ approaches its bound varies by configuration; round markers in Fig.~\ref{fig:DownloadCapacity_vs_N} indicate the 75\% capacity threshold. Higher CubeSat altitudes reach the threshold with fewer \gls{ntn} relays. Comparing technologies, Ku-band and Sub-THz NG saturate earlier than current Sub-THz hardware, which remains power-limited—underscoring the need for higher-power Sub-THz sources. Notably, the threshold of diminishing returns often occurs below $N_{\text{S}}=20$, where the marginal capacity gains from additional relays may not justify the increased cost.
\begin{figure}[!b]
    \centering
    \includegraphics[width=1\linewidth]{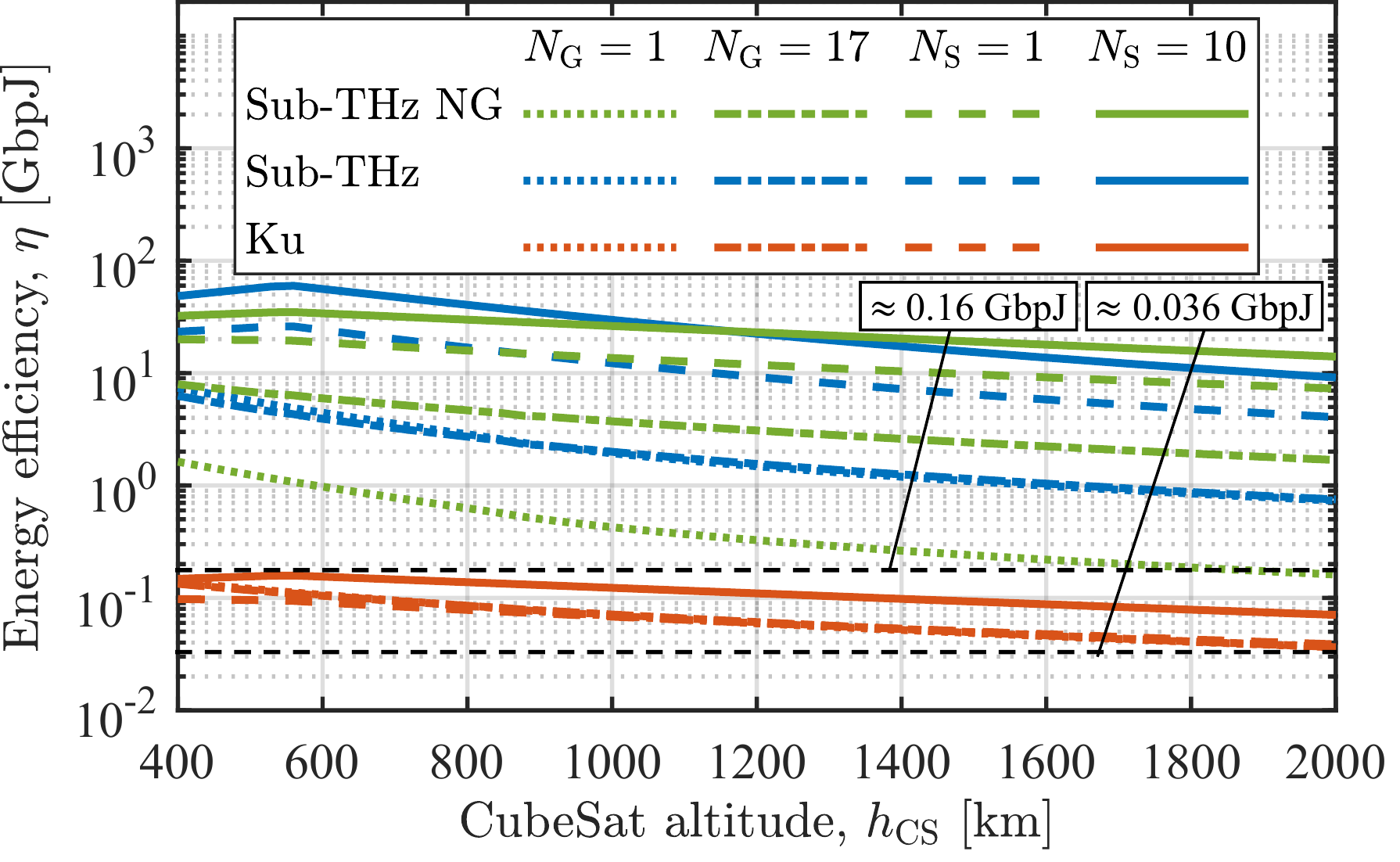}
    \caption{Energy efficiency of the CubeSat links.}
    \vspace{-5mm}
    \label{fig:ee}
\end{figure}

\subsection{Energy efficiency}\label{subsec:ee_num}
To wrap up the results, we analyze the energy efficiency $\eta$ across technologies and architectures in Fig.~\ref{fig:ee}. For \gls{gs} architectures, $\eta$ decreases monotonically with altitude because the gains in total download capacity $\Gamma$ fail to offset the increased link distance and contact time trade-offs described in~\eqref{eq:energyeff}. In contrast, \gls{ntn} architectures achieve peak energy efficiency at approximately $550$~km, directly coinciding with the altitude of the relay nodes. Ku-band energy efficiency remains low, between $0.036$ and $0.16$~GbpJ, while Sub-THz technologies achieve orders-of-magnitude higher efficiency. This advantage is driven by vast available bandwidths, enabling superior total download capacity despite using transmit powers $100$ and $20$ times lower than Ku-band for current and next-generation cases, respectively. Notably, increasing Sub-THz transmit power slightly reduces energy efficiency but allows the link to escape the power-limited regime. This trade-off is justified by the substantial throughput gains—as seen in Fig.~\ref{fig:DownloadCapacity}—provided by next-generation hardware.

\section{Conclusions}\label{sec:conclusions}
As the commercial space economy grows, 6G and beyond \glspl{ntn} offer a vital paradigm shift to overcome the bandwidth constraints and \gls{gs} congestion facing emerging ``space users". In this study, we evaluated this new concept via a comprehensive mathematical framework that incorporates satellite orbital dynamics, mutual mobility, and radio link characteristics. We then quantified the key performance indicators for such a system, including the contact probability, capacity, daily download volume, and energy efficiency. To ensure accuracy, our analytical results were also cross-verified via computer simulations. We finally compared the performance boundaries for the considered NTN-relay-assisted deployment scenarios with state-of-the-art Ku-band \glspl{gds} (a representative baseline~\cite{kodheli2021satellite}).

Our key findings from this study are:
\begin{enumerate}[leftmargin=5mm] 
\item \Gls{ntn}-based relays provide \emph{substantial improvements in service availability} compared to ground relays. Particularly, the probability of contact in Fig.~\ref{fig:linkdistance_comparison} for as few as 3 \gls{ntn} relays is already much higher than the one with as many as 17 ground relays. 
\item \emph{Continuous 24/7 coverage} is achievable with as few as 8--10 \gls{ntn} satellites per orbit -- in contrast to only up to $40\%$ contact probability with roughly twice as many GSs for CubeSats at altitudes under $1000$~km (hence, over $60\%$ of time, the GS-connected CubeSat is unavailable).
\item The utilization of \gls{ntn}-based relays also leads to \emph{up to a tenfold increase} in the downlink capacity and the daily download volume (as illustrated in Fig.~\ref{fig:cap_vs_t}--Fig.~\ref{fig:DownloadCapacity}).
\item Assuming continuous beam alignment, the use of sub-THz radio links, especially, the forthcoming sub-THz NG-class equipment may lead to \emph{up to $40$x increase} in the total download volume of data per day compared to state-of-the-art Ku radio in similar conditions\footnote{Imperfect beam alignment is expected to lead to performance gains of sub-THz over Ku becoming lower than $40$x but still staying substantial.}.
\item While a higher number of \gls{ntn} relays naturally increases capacity, the system reaches diminishing extra benefits at approximately 20 nodes per orbit, bounding the benefit of further densification (as best illustrated in Fig.~\ref{fig:DownloadCapacity_vs_N}).
\item Last but not least, the use of \gls{ntn}-based space relays also notably improves the energy efficiency of the data exchange with the CubeSat (as summarized in Fig.~\ref{fig:ee}). 
    \end{enumerate}

The shift toward mmWave- and sub-THz-enabled \glspl{ntn} marks a transformative milestone for space exploration and commercialization. By dismantling the ground-based download bottleneck, the concept evaluated in this article provides a scalable, multi-gigabit backbone for next-generation data-intensive missions. As 6G matures, high-frequency relay constellations will serve as the foundational infrastructure for a connected space economy. Ultimately, combining the benefits of mmWave and sub-THz systems with the ubiquitous coverage of \gls{leo} satellite communication networks transforms emerging space users (CubeSats, space telescopes, orbital stations, etc.) from \emph{only sporadically-connected individual devices} into \emph{a continuously-available integral part} of 6G+ heterogeneous non-terrestrial networks of tomorrow.

\appendices{}
\section{CDF of the distance to an NTN satellite}\label{app:CDF_derivation}
The CDF of the distance between the CubeSat and an NTN satellite is defined as $F_{d_{\text{NTN}}}(d) = P(d_{\text{NTN}}(t) \leq d)$. We incorporate \eqref{eq:d_ntn} in the expression above to obtain:
\begin{equation}
\begin{split}
    F&_{d_{\text{NTN}}}(d) =P\left( \cos(\Delta \omega t) \hspace{-1mm} \geq \hspace{-1mm} \sigma \right) P\hspace{-1mm}\left( 0 < t < \frac{T_{\text{NTN}}}{2} \right) +\\
&P\left( \cos(\Delta \omega(T_{\text{NTN}} - t)) \hspace{-1mm}\geq \hspace{-1mm}
\sigma\right) P\hspace{-1mm}\left( \frac{T_{\text{NTN}}}{2} < t < T_{\text{NTN}} \right),
\end{split}
\end{equation}
where $\Delta \omega=|\omega_{\mathrm{CS}} - \omega_{\mathrm{NTN}}|$, and:
\begin{equation}
    \sigma = \frac{R_{\mathrm{CS}}^2 + R_{\mathrm{NTN}}^2 - d^2}
{2 R_{\mathrm{CS}}R_{\mathrm{NTN}}}.
\end{equation}

Therefore, solving the inner inequalities for t:
\begin{equation}
    \begin{split}
        F_{d_{\text{NTN}}}&(d) = \frac{1}{2} P\left( 0 \leq t \leq \frac{1}{\Delta \omega} \cos^{-1} \left(\sigma\right) \right)  \\
& + \frac{1}{2} P\left( T_{\text{NTN}} - \frac{1}{\Delta \omega} \cos^{-1} \left( 
\sigma\right) \leq t \leq T_{\text{NTN}} \right).
    \end{split}
\end{equation}

Here, we incorporate the assumption that $t\sim U[0, T_{\text{NTN}}]$ to obtain:
\begin{equation}
    F_{d_{\text{NTN}}}(d) = \frac{2}{T_{\text{NTN}} \Delta \omega} \cos^{-1} \left(\sigma\right).
\end{equation}

Finally, incorporating \eqref{eq:Tdntn} we obtain the final expression of the CDF when $d\in(d_{\text{NTN}}^{\min}, \max\{d_{\text{NTN}}^{\text{B}}, d_{\text{NTN}}^{\max}\})$:
\begin{equation}
    F_{d_{\text{NTN}}}(d) = \frac{N}{\pi} \cos^{-1} \left( 
\frac{R_{\mathrm{CS}}^2 + R_{\mathrm{NTN}}^2 - d^2}
{2 R_{\mathrm{CS}}R_{\mathrm{NTN}}}\right).
\end{equation}

\balance
\bibliographystyle{IEEEtran}
\bibliography{bibliography}

\end{document}